\documentclass[twocolumn, times, tighten]{aastex61}
\usepackage{textcomp}
\usepackage{amsmath}
\usepackage{color}
\usepackage{amssymb}
\usepackage{hyperref}

\newcommand{\MgII}{Mg\,{\sc ii}}
\newcommand{\HeII}{He\,{\sc ii}}
\newcommand{\CIV}{C\,{\sc iv}}

\newcommand{\OIII}{O\,{\sc iii]}}
\newcommand{\Hbeta}{H$\beta$}

\shorttitle{Sun et al.}

\begin{document}
\revised{\bf Draft: \today}
\title{The Sloan Digital Sky Survey Reverberation Mapping Project: 
the \CIV\ Blueshift, Its Variability, and Its Dependence Upon Quasar 
Properties}

\author[0000-0002-0771-2153]{Mouyuan Sun}
\affiliation{CAS Key Laboratory for Research in Galaxies and Cosmology, 
Department of Astronomy, University of Science and Technology of China, Hefei 
230026, China; ericsun@ustc.edu.cn; xuey@ustc.edu.cn}
\affiliation{School of Astronomy and Space Science, University of Science 
and Technology of China, Hefei 230026, China}

\author[0000-0002-1935-8104]{Yongquan Xue}
\affiliation{CAS Key Laboratory for Research in Galaxies and Cosmology, 
Department of Astronomy, University of Science and Technology of China, Hefei 
230026, China; ericsun@ustc.edu.cn; xuey@ustc.edu.cn}
\affiliation{School of Astronomy and Space Science, University of Science 
and Technology of China, Hefei 230026, China}

\author{Gordon T. Richards}
\affiliation{Department of Physics, Drexel University, 3141 Chestnut St., Philadelphia, PA 19104, USA}

\author{Jonathan R. Trump}
\affiliation{Department of Physics, University of Connecticut, Storrs, CT 06269, USA}

\author{Yue Shen}
\altaffiliation{Alfred P. Sloan Research Fellow}
\affiliation{Department of Astronomy, University of Illinois at Urbana-Champaign, Urbana, IL 61801, USA}
\affiliation{National Center for Supercomputing Applications, University of Illinois at Urbana-Champaign, 
Urbana, IL 61801, USA}

\author{W. N. Brandt}
\affiliation{Department of Astronomy \& Astrophysics, 525 Davey Lab, The Pennsylvania State
  University, University Park, PA 16802, USA}
\affiliation{Institute for Gravitation and the Cosmos, 525 Davey Lab, The Pennsylvania State
  University, University Park, PA 16802, USA}
\affiliation{Department of Physics, 104 Davey Lab, The Pennsylvania State University, University Park, PA 
16802, USA}

\author{D. P. Schneider}
\affiliation{Department of Astronomy \& Astrophysics, 525 Davey Lab, The Pennsylvania State
  University, University Park, PA 16802, USA}
\affiliation{Institute for Gravitation and the Cosmos, 525 Davey Lab, The Pennsylvania State
  University, University Park, PA 16802, USA}

\begin{abstract}
We use the multi-epoch spectra of 362 quasars from the Sloan Digital Sky Survey Reverberation Mapping 
project to investigate the dependence of the blueshift of \CIV \ relative to \MgII \ on quasar 
properties. We confirm that high-blueshift sources tend to have low \CIV \ equivalent widths (EWs), 
and that the low-EW sources span a range of blueshift. Other high-ionization lines, such as \HeII, 
also show similar blueshift properties. The ratio of the line width (measured as both 
the full-width at half maximum and the velocity dispersion) of \CIV\ to that of \MgII\ 
increases with blueshift. Quasar variability might enhance the connection between the \CIV \ 
blueshift and quasar properties (e.g., EW). The variability of the \MgII\ line center 
(i.e., the wavelength that bisects the cumulative line flux) 
increases with blueshift. In contrast, the \CIV\ line center shows weaker variability 
at the extreme blueshifts. Quasars with the high-blueshift CIV lines tend to have less 
variable continuum emission, when controlling for EW, luminosity, and redshift. Our results 
support the scenario that high-blueshift sources tend to have large Eddington ratios. 
\end{abstract}


\section{Introduction}
\label{sect:intro}
Broad emission lines (hereafter BELs) from the ``broad line region'' (BLR) 
are unambiguous features in quasar spectra. These BELs can be divided into two 
main categories based upon their ionization potential: high-ionization BELs (e.g., 
\CIV, \HeII) with ionization energy $E_{\rm{ion}}\gtrsim 50\ \rm{eV}$, and 
low-ionization BELs (e.g., H$\alpha$, \Hbeta, \MgII) with ionization energy 
$E_{\rm{ion}}\lesssim 50\ \rm{eV}$. Compared with low-ionization BELs, 
high-ionization BELs are believed to be produced closer to the central supermassive 
black holes (SMBHs). Interestingly, high-ionization BELs, such as \CIV, are often 
significantly shifted blueward with respect to their low-ionization counterparts 
\citep[e.g.,][]{Gaskell1982, Wilkes1986, Corbin1990, Sulentic2000a, Sulentic2007, 
Baskin2005, Richards2002, Richards2011, Wang2011, Denney2012, Shen2008, Shen2012, 
Coatman2016, Coatman2017}. The blueshift velocity can be as large as $\sim 8000\ 
\rm{km\ s^{-1}}$ \citep[e.g.,][]{Luo2015}. This result raises several important questions, 
including the origin of the blueshift, its effect on the estimation of the mass of the 
central SMBHs ($M_{\rm{BH}}$), and its role in quasar unification.  

The blueshift is often attributed to accretion-disk winds \citep[e.g.,][]{Gaskell1982, 
Murray1997, Leighly2004, Leighly2007, Richards2011, Denney2012, Chajet2013}. 
Such winds can disturb the velocity field of the BLR material and therefore can affect 
the line profiles. An alternative possibility is that the blueshift is due to the scattering 
between the inflowing gases and the BEL photons \citep{Gaskell2009}. There are also 
speculations that the blueshift arises due to radiative transfer effects \citep{Richards2002}. 

Different physical drivers can be assessed using correlations of blueshift with quasar 
properties. For instance, it is likely that high-blueshift quasars favor 
distinct regions of quasar parameter space or the quasar Eigenvector 1 sequence 
\citep{Boroson1992, Sulentic2000b, Sulentic2017, Dong2009, Runnoe2014, Shen2014}, e.g., high 
Eddington ratios $\lambda_{\rm Edd}$ \citep{Baskin2005, Coatman2016}, soft (i.e., with weak 
X-ray emission) spectral energy distributions \citep[SEDs; see e.g.,][]{Leighly2004, Richards2011, 
Luo2015}, low inclination angles \citep{Denney2012} and/or low variability. 

In addition to the origin of the blueshift and its correlation with quasar properties, 
it is also important to consider the effects of the blueshift on $M_{\rm{BH}}$ 
estimation. As noted by \cite{Richards2011}, the widely adopted single-epoch 
black-hole mass estimators (e.g., \citealt{Vestergaard2006, Vestergaard2009, Shen2011}; for 
a recent review, see \citealt{Shen2013a}) are derived using a low-blueshift reverberation-mapped 
quasar sample. There are two reasons why such estimators might not be valid for high-blueshift 
quasars. First, as mentioned before, the observed line profiles are likely due to a mixture of 
virial (i.e., dominated by the 
gravitational potential of the central SMBHs) and non-virial motions. Second, there 
are indications that the empirical BLR radius-quasar optical luminosity relation for 
\Hbeta \ depends on quasar SEDs \citep{Kilerci2015} and/or Eddington ratios \citep{Du2016}. 
If the high-blueshift quasars indeed occupy a distinct region of quasar parameter 
space, the current radius-luminosity relation could be invalid for those quasars. 

The blueshift was not the first BEL feature showing significant changes across 
the quasar distribution. Rather that was the well-known Baldwin Effect \citep{Baldwin1977}, 
which represents an anti-correlation between the \CIV\ EW and luminosity 
\citep[e.g.,][]{Dietrich2002, Wu2009}. Weak BEL quasars also tend to show large \CIV\ blueshifts 
\citep[e.g.,][]{Marziani1996, Marziani2016, Richards2011, Plotkin2015}. Hence, there could 
be possible connections between the \CIV\ blueshift and EW. It is now well-established that 
{\em both} the EW and blueshift are needed to minimally characterize the range of properties 
exhibited by \CIV\ \citep{Sulentic2007, Richards2011, Marziani2016}. 

In this work, we explore the high-ionization BEL blueshift phenomenon taking advantage 
of the first $32$ epochs of spectra from the Sloan Digital Sky Survey Reverberation Mapping 
project \citep[SDSS-RM; for a technical overview, see][]{Shen2015}. Compared with previous 
works, the SDSS-RM project provides 
a high S/N composite spectrum for each of the $849$ quasars, allowing us to measure 
accurately the blueshift for both strong and weak emission lines. By analyzing the spectra 
epoch by epoch, we can also measure the variability properties of the blueshift. Finally, 
understanding the blueshift properties of the SDSS-RM sample is also 
crucial for the project since one of its main goals is to provide unbiased 
$M_{\rm{BH}}$ estimators for a wide variety of quasars.  

This paper is formatted as follows. In Section~\ref{sect:measure}, we discuss our spectral 
fitting procedures, and our measurements of quasar properties. In Section~\ref{sect:result_cp} 
we show our analysis of the high S/N composite spectra. In Section~\ref{sect:result_sp}, 
we present measurements of the variability of the blueshift. In Section~\ref{sect:dis}, we 
discuss the implication of our results. A summary of our work appears in Section~\ref{sect:sum}. 
We adopt a flat $\Lambda$CDM cosmology with $h_0=0.7$ and $\Omega_{\rm{M}} = 0.3$. 
Throughout this work, the wavelengths of quasar features always refer to the rest-frame, 
unless otherwise specified.

\section{Spectral measurement}
\label{sect:measure}
The SDSS-RM project is an ancillary program within the SDSS-III \citep{Eisenstein2011} BOSS survey 
\citep{Dawson2013} using a dedicated $2.5$ m telescope at Apache Point Observatory \citep{Gunn2006}. 
The spectrograph has a wavelength range of $3650$--$10400$ $\rm{\AA}$ with a spectral resolution of 
$R\sim 2000$ \citep{Smee2013}. Each of the $32$ epochs has a typical exposure time of $2$ hours. 
The spectra were pipeline-processed \citep{Bolton2012} and were flux calibrated via a custom scheme 
\citep{Shen2015}. The SDSS-RM sample consists of $849$ BEL quasars. We only select $1.48<z<2.6$ 
sources for which \MgII \ and \CIV \ were both covered in the BOSS spectra. We focus on $29$ (three 
of the $32$ epochs are discarded due to low S/N) epochs of the SDSS-RM \citep{Shen2015} spectra 
and the resulting high S/N composite spectra. As mentioned in \cite{Shen2015} and \cite{Sun2015}, 
there are spectra with flux anomalies.\footnote{As noted by \cite{Shen2015}, such 
spectra might be obtained due to the dropping of the fiber during spectroscopic exposures.} An 
epoch was identified as an outlier if its flux is more than $1$ magnitude away from the median of 
all epochs \citep{Sun2015}. These outliers are rejected. Below we explain our spectral fitting 
approach. 

\subsection{Spectral fitting}
\label{sect:spfit}
\subsubsection{Continuum fitting}
\label{sect:ctfit}
Our spectral-fitting approach is similar to that of \cite{Trump2009} and \cite{Shen2011}. 
For each spectrum, we first fit a double power-law continuum (i.e., $f_{\lambda}=A_{1} 
\lambda^{\beta_{1}}$ if $\lambda<2000\ \rm{\AA}$; $f_{\lambda}=A_{2} 
\lambda^{\beta_{2}}$ if $\lambda>2000\ \rm{\AA}$) plus a broadened iron template 
\citep{ironuv} to the following relatively emission line-free wavelength ranges, $1445\ 
\mathrm{\AA}<\lambda <1465\ \mathrm{\AA}$, $1700\ \mathrm{\AA}<\lambda<1705\ 
\mathrm{\AA}$, $2200\ \mathrm{\AA}<\lambda<2700\ \mathrm{\AA}$, and $2900\ 
\mathrm{\AA}<\lambda < 3088\ \mathrm{\AA}$. During the continuum and the subsequent 
emission-line fitting, we rejected data points that are $3\sigma$ below the $30$-pixel 
boxcar-smoothed spectrum. The purpose of this procedure is to reduce the effects of 
narrow absorption lines. We performed an iterative $\chi^2$ minimization\footnote{We 
use \textit{kmpfit}, a Python version of the least squares fit routine, to perform 
our fitting. This routine is available as a part of the Kapteyn package, which 
can be downloaded from \url{http://www.astro.rug.nl/software/kapteyn/}.} to optimize 
the fits. The continuum and iron best fit is then subtracted from 
the spectrum. We then fit the resulting line spectrum with several Gaussian functions. 
In the following sections, we present our modeling procedures for \MgII \ and \CIV \ 
(and \HeII\ $\lambda$1640).

\subsubsection{Line fitting}
\label{sect:linefit}
\MgII: We fit the continuum- and iron-subtracted flux in the wavelength range of $2700\ 
\mathrm{\AA}<\lambda<2900\ \mathrm{\AA}$ with three Gaussian functions, each with an 
unconstrained full-width at half maximum (FWHM), i.e., we do not consider narrow \MgII \ 
subtraction. We calculate the \MgII\ line-profile properties from the overall line profile 
which is the summation of the multiple best-fit Gaussian functions. Any Gaussian function 
with the ratio of its flux to the total line flux $<0.05$ is ignored. 

\CIV: We adopted different line-modeling procedures for the composite and single-epoch 
spectra. 

For the composite data, the pseudo-continuum subtracted spectrum from $1500\ 
\mathrm{\AA}<\lambda <1700\ \mathrm{\AA}$ was modeled with six Gaussian functions: 
two Gaussians for \CIV, two Gaussians for \HeII\ $\lambda 1640$, and the remaining two 
Gaussians for \OIII\ $\lambda 1663$. Therefore, the $1600\ \mathrm{\AA}$ 
feature of the composite data is modeled by the superposition of the red tail of 
\CIV\ and a broad \HeII . This approach is similar to some previous studies 
\citep[e.g.,][]{Fine2010, Marziani2010}. Similar to that of \MgII, we do not set 
limits on the FWHMs of the Gaussian functions. 

For the single-epoch spectra, we only considered the following wavelength range: 
$1500\ \mathrm{\AA}<\lambda <1600\ \mathrm{\AA}$. The resulting pseudo-continuum 
subtracted spectrum was fitted with two Gaussian functions. We do not model either 
\HeII\ or \OIII\ since the main purpose of the single-epoch spectral fitting is to 
constrain the variability of line properties. The measurement errors of \HeII\ or 
\OIII\ are relatively large since the two lines are weak. Therefore, it is nontrivial 
to constrain reliably the intrinsic variability of their line properties. 

Similar to \MgII, any Gaussian function with the ratio of its flux to the total flux 
$<0.05$ is removed from consideration when calculating the line-profile properties. 

\begin{figure*}
\epsscale{1.2}
\plotone{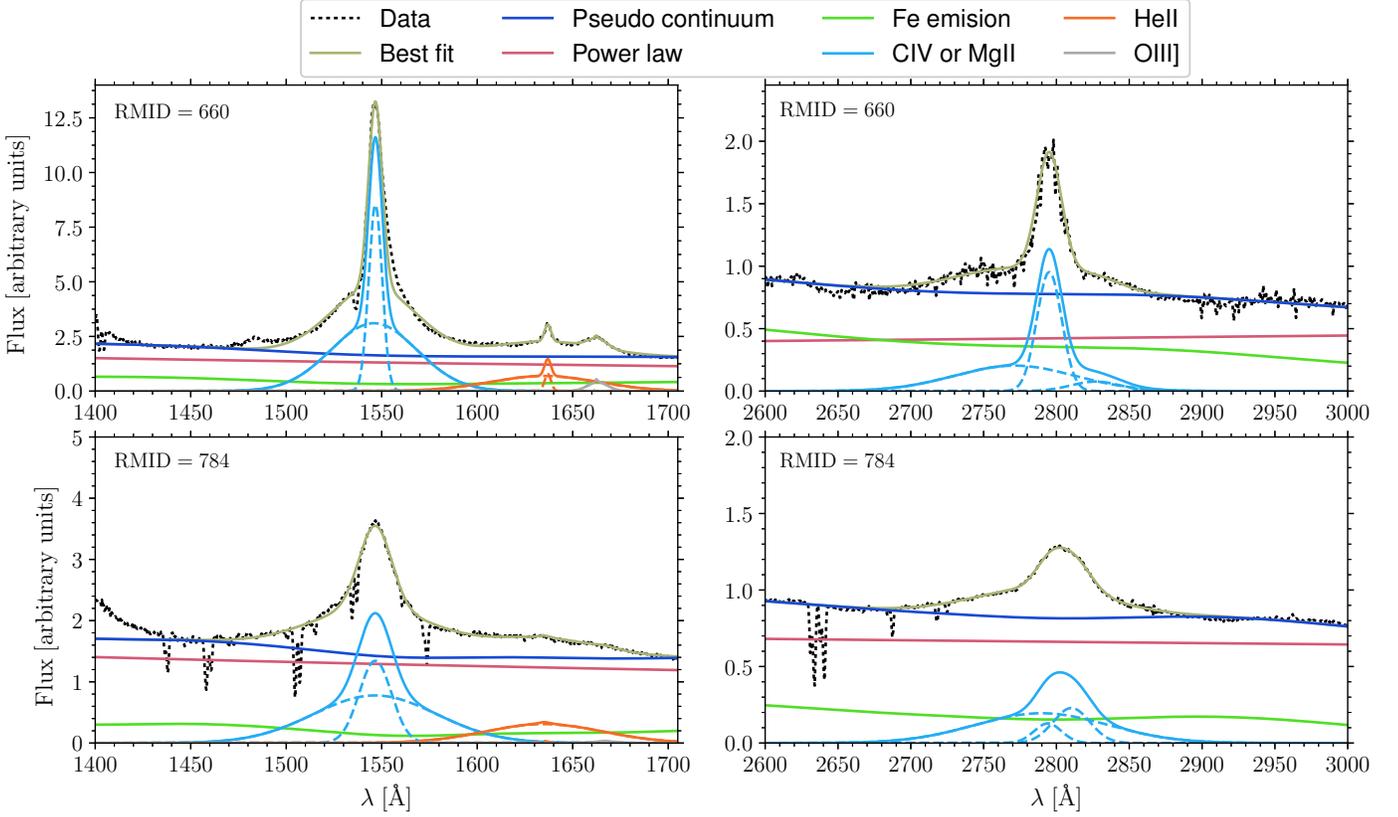}
\caption{Examples of multi-component fits to the composite spectra. The upper and lower 
panels are for RMID = $660$ and $784$, respectively. The left and 
right panels are for \CIV\ (with \HeII\ and \OIII) and \MgII, respectively. }
\label{fig:fit_example}
\end{figure*}

In Figure~\ref{fig:fit_example}, we present examples of our fits to the 
high S/N composite spectra of RMID\footnote{RMID is the index of sources in 
the SDSS-RM catalog \citep[see Table 1 of][]{Shen2015}.} =660 and RMID=784. 

\subsubsection{Uncertainty estimation}
\label{sect:error}
We adopted a Monte Carlo approach to estimate the uncertainties of the spectral-fitting 
parameters. A total of $100$ ($50$ for single-epoch spectra) mock spectra were synthesized, 
where the flux in each wavelength pixel was generated by adding the flux density noise to 
the best-fit models. We then fit the mock spectra following the same fitting 
recipe. The uncertainties are estimated from the statistical dispersion of the best-fit models 
of the mock spectra. The statistical dispersion was estimated by $0.74\mathrm{IQR}(x)$ 
where IQR($x$) is the interquartile range (IQR) of the variable $x$. The constant $0.74$ 
normalizes the IQR to be equivalent to the standard deviation of a Gaussian distribution. 
Unlike the standard deviation, the IQR is robust against outliers or tails in the distribution. 

Following \cite{Shen2013b}, we justified our uncertainty estimation by exploring the 
distributions of quasar properties between close (i.e., rest-frame time interval 
$< 2$ days) pairs. We then compared these distributions with the expected ones from 
the measurement errors. Our Monte Carlo approach underestimated the true uncertainties, 
so we enlarged the uncertainties by a constant factor until the expected distributions from 
the measurement errors matched the observed close-pair distributions. The constant factor 
varies from $\sim 1.2$ to $\sim 1.7$, depending on the physical quantities we are interested 
in. In the following analyses, we will scale our uncertainties up by the constant factor.

\subsection{Emission-line properties}
\label{sect:emlp}
\begin{figure}
\epsscale{1.2}
\plotone{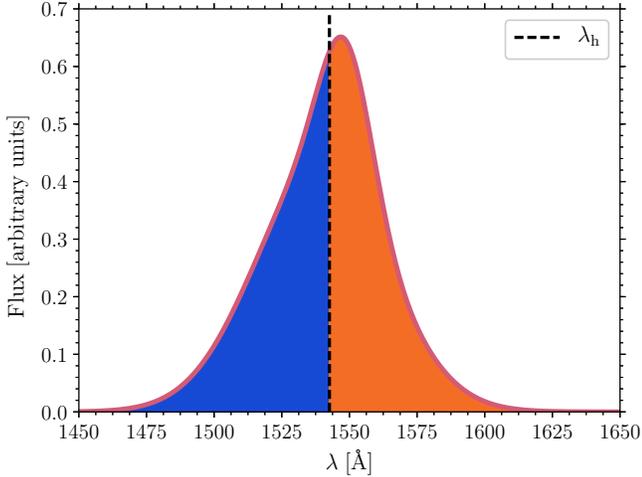}
\caption{An illustration of the definition of $\lambda_{\rm{h}}$. That is, the total flux 
of the wavelengths shortward (i.e.,$\lambda<\lambda_{\rm{h}}$) equals to that of the wavelengths longward 
(i.e., $\lambda>\lambda_{\rm{h}}$). }
\label{fig:ct_def}
\end{figure}

We calculated the following parameters of the emission-line properties. All line measurements 
are from the total line profile, which is the sum of the multiple Gaussians (excluding Gaussian 
components that contribute less than $5\%$ of the total flux). 
\begin{itemize}
\item[1.] The shift velocity, $V_{\rm{shift}}$ ($V_{\rm{shift, se}}$ for the single-epoch 
data), is defined as $c\times (\lambda_{\rm{h}} - \lambda_{\rm{va}})/\lambda_{\rm{va}}$, where 
$c$, $\lambda_{\rm{va}}$, and $\lambda_{\rm{h}}$ are the speed of light, the central wavelength 
of the emission line in vacuum, and the line center. The latter is defined 
as the wavelength that bisects the cumulative total line flux \citep{Coatman2017}. 
Figure~\ref{fig:ct_def} presents an illustration of the definition of 
$\lambda_{\rm{h}}$. 

\item[2.] The offset of \CIV \ for the coadded spectrum is, $V_{\mathrm{off,CIV}}=
V_{\rm{shift,CIV}}-V_{\rm{shift,MgII}}$.\footnote{We adopted this definition because, for 
the redshift ranges considered here, \MgII\ is the best practical redshift estimator 
\citep{Shen2016}.} That is, negative values indicate blueshift toward 
the observer (i.e., ``outflows''). We define the single-epoch offset velocity 
as $V_{\mathrm{off,se}} =V_{\rm{shift,CIV,se}}-V_{\rm{shift,MgII}}$. The observed variations 
of $V_{\mathrm{off,se}}$ are due to the line-shift variability of \CIV. The offsets of other 
lines are defined in a similar way. Our definition of the offset of \CIV \ might be an underestimation 
of the true value because \MgII\ lines also show offsets with respect to \Hbeta\ \citep{Marziani2013} 
or the host galaxy \citep[e.g.,][]{Shen2016} by a median blueshift velocity of $65\ \mathrm{km\ 
s^{-1}}$ with an intrinsic scatter of $\sim 200\ \mathrm{km\ s^{-1}}$.

\item[3.] The emission-line velocity width can be measured by FWHM or dispersion ($\sigma$) 
of the profile. Compared with FWHM, $\sigma$ is more sensitive to the wing of the 
emission line. 

\item[4.] The emission-line shape is defined as $D=\rm{FWHM}/\sigma$. For a perfect Gaussian 
profile, $D=2.354$. Larger values of $D$ indicate that the profiles are more ``boxy''. 

\item[5.] The equivalent width is calculated using $\rm{EW}=\int_{\lambda_{\rm{va}}-200\ 
\mathrm{\AA}}^{\lambda_{\rm{va}} +200\ \rm{\AA}} 
\frac{f_{\rm{line}}(\lambda)}{f_{\rm{cont}}(\lambda)}\mathrm{d}\lambda$, where 
$f_{\rm{line}}(\lambda)$ and $f_{\rm{cont}}(\lambda)$ are both obtained from our spectral fitting 
results. 
\end{itemize} 

We measured the $1350\ \rm{\AA}$ and the $3000\ \rm{\AA}$ continuum luminosities (hereafter 
$L1350$ and $L3000$, respectively) from the best-fit double power-law component. 
We adopted the $1350\ \rm{\AA}$ continuum luminosity multiplied by a monochromatic 
bolometric correction of $5$ \citep{Richards2006} as an estimator of the bolometric luminosity, 
$L_{\rm{Bol}}$.\footnote{Our results do not critically depend on $L_{\rm{Bol}}$. Therefore, our 
conclusions do not change if we instead adopt luminosity-dependent bolometric correction factors 
\citep[e.g.,][]{Lusso2012, Krawczyk2013}.} 

We adopt the \MgII \ virial estimator to measure $M_{\rm{BH}}$ \citep[see 
Eq.~(8) of][]{Shen2011}. The Eddington ratio is $\lambda_{\rm{Edd}}= 
L_{\rm{Bol}}/(1.26\times 10^{38} M_{\rm{BH}}/M_{\odot}\ \rm{erg\ s^{-1}})$. 

\subsection{Sample properties}
\label{sect:sample}
\begin{figure}
\epsscale{1.2}
\plotone{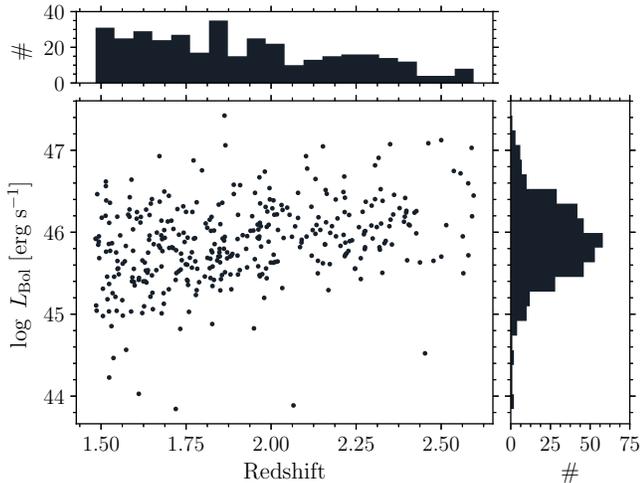}
\caption{Distribution of our sample  in the $L_{\rm{Bol}}$-redshift plane. Our sample 
consists of $362$ sources with $1.48<z<2.6$. The quasar luminosity range spans two 
orders of magnitude. }
\label{fig:lbol_z}
\end{figure}

We flagged our fits to the composite spectra by visual inspection. The spectra of 
quasars for which reliable emission-line parameters could not be estimated 
are rejected. These spectra contain strong broad absorption lines, or have multiple 
absorption features around the line centers of \MgII \ or \CIV. Our final sample consists 
of $362$ sources. In Fig.~\ref{fig:lbol_z}, we present the distribution of $L_{\rm{Bol}}$ 
as a function of redshift for this sample. The quasar luminosity, $M_{\mathrm{BH}}$ 
and $\lambda_{\mathrm{Edd}}$ ranges span two orders of magnitude, and therefore 
it is suitable to explore the dependencies of the \CIV\ blueshift upon quasar properties. 

\section{The composite spectra}
\label{sect:result_cp}
We are now in a position to explore $V_{\mathrm{off,CIV}}$ as a function of emission 
line properties. 

\subsection{The blueshift and EW}
\label{sect:voffew}
\begin{figure}
\epsscale{1.2}
\plotone{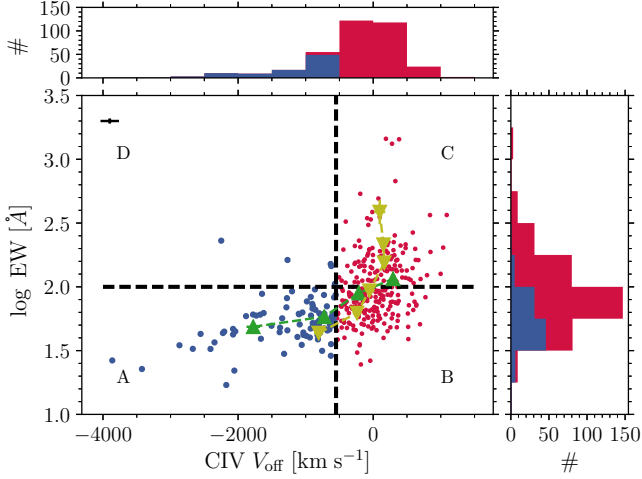}
\caption{Distribution of our sample in the \CIV \ offset velocity-EW plane. Sources with offset velocity 
$<-550\ \rm{km\ s^{-1}}$ are highlighted by blue colors. The green (yellow) triangles represent 
the mean $\log\ \rm{EW}$ (\CIV \ $V_{\rm{off}}$) in each \CIV \ $V_{\rm{off}}$ ($\log\ \rm{EW}$) bin. 
It is clear that the high-blueshift quasars tend to have small EWs. However, the scatter of the correlation 
is not negligible. We defined three samples A, B, and C according to the distribution (black dashed lines; 
see texts). Similar to the results of \cite{Richards2011}, our sources tend to avoid the high-blueshift and 
high-EW (i.e., ``D'') space. The black cross indicates the typical uncertainties of the \CIV \ offset velocity 
and EW. }
\label{fig:ew_vof}
\end{figure}

Figure~\ref{fig:ew_vof} shows the distribution of our sample in the \CIV \ EW-offset velocity 
parameter space. The distribution of the \CIV \ offset velocity is not symmetric, with a long 
tail of blueshifted velocities. For instance, $\sim 5\%$ of sources have $V_{\mathrm{off,CIV}}> 
550\ \rm{km\ s^{-1}}$ while $\sim 24\%$ of sources (highlighted as blue dots) have 
$V_{\mathrm{off,CIV}} <-550\ \rm{km\ s^{-1}}$. 

We also binned the sources according to the \CIV \ EW or the offset velocity. Consistent with 
previous works \citep[e.g.,][]{Richards2011, Luo2015}, the sources with extreme \CIV \ blueshifts 
tend to have weak \CIV. However, weak \CIV \ is an insufficient condition for a quasar to have a 
strong \CIV\ blueshift. We will further discuss the connection between the \CIV\ blueshift and 
EW in Section~\ref{sect:dis}. Our sources can be divided into three subsamples that will be 
used in some of the subsequent analyses: 
\begin{itemize}
\item[1.] Sample A: the ``blueshift'' sub-sample, i.e., sources with offset velocities $<-550\ 
\rm{km\ s^{-1}}$ and $\log \rm{EW}<2.0$ (we selected this limit because $95\%$ of sources 
with offset velocities $<-550\ \rm{km\ s^{-1}}$ satisfy this limit). There are $84$ sources in 
this sample. 

\item[2.] Sample B: sources with offset velocities $>-550\ \rm{km\ s^{-1}}$(i.e., weak or no blueshift) 
and $\log \rm{EW}<2.0$ (i.e., weak \CIV). $154$ sources belong to this sample. 

\item[3.] Sample C: sources with offset velocities $>-550\ \rm{km\ s^{-1}}$ (i.e., weak or no 
blueshift) and $\log \rm{EW}>2.0$ (i.e., strong \CIV). This sample consists of $118$ sources. 
\end{itemize}
There are only $6$ sources in the high-blueshift and high-EW space. 

\begin{figure}
\epsscale{1.2}
\plotone{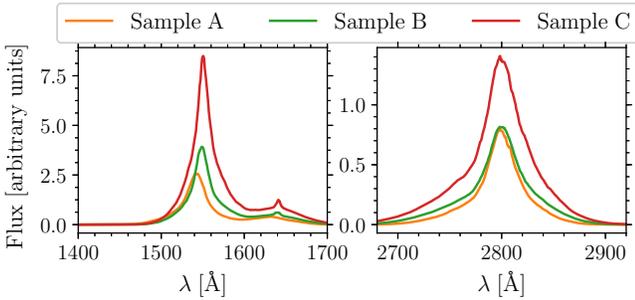}
\caption{Composite spectra from \CIV, \HeII\ and \MgII\ for the three regions of the \CIV \ offset velocity-EW 
plane (see Figure~\ref{fig:ew_vof}). The composite spectra are normalized to the best-fitting $2400\ \mathrm{\AA}$ 
continuum. We adopted \MgII\ as the redshift estimator. It is evident that both \CIV\ and \HeII\ show blueshift 
with respect to \MgII. The \CIV\ shape parameter $D$ of Sample A is larger than that of Sample B or C. Therefore, 
the \CIV\ line profile of Sample A is more boxy than Sample B or C. }
\label{fig:stack}
\end{figure}

We have constructed composite spectra of \MgII , \HeII\ and \CIV\ for samples A, B, and C. The procedures 
to stack individual spectra into a composite spectrum are as follows. First, we normalize each individual 
spectrum by its best-fitting $2400\ \mathrm{\AA}$ continuum flux. Second, for each wavelength, we take 
the median flux from the best-fitting line profiles of the normalized individual spectra. Third, we shift the 
wavelength to ensure that $V_{\mathrm{shift,MgII}}=0$, i.e., we adopt \MgII\ as the redshift estimator. In 
Figure~\ref{fig:stack}, we show the three composite spectra of \MgII , \HeII\ and \CIV . Both \CIV\ and 
\HeII\ show blueshift with respect to \MgII . The line shapes of \CIV\ and \HeII\ also changes with the 
\CIV\ blueshift. 

\begin{figure}
\epsscale{1.2}
\plotone{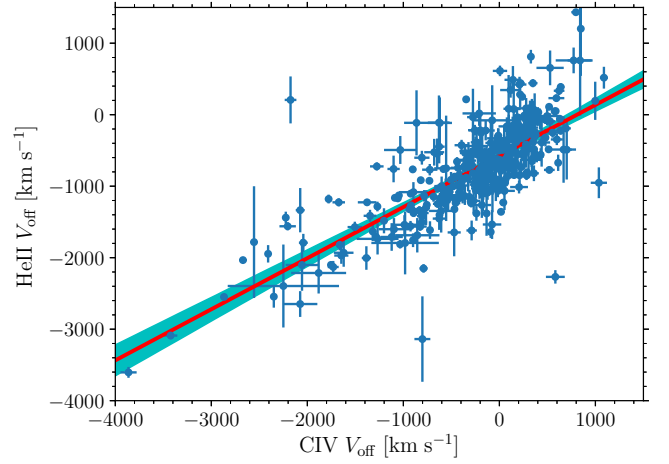}
\caption{The \HeII\ offset as a function of the \CIV \ offset (both are relative to \MgII). The red 
solid line and the shadowed region represent the best-fit relation and its $1\sigma$ confidence 
band. The best-fit relation is $y=(0.69\pm 0.02)x-(583\pm 20)\ \mathrm{km\ s^{-1}}$, where $x$ 
and $y$ correspond to the blueshifts of \CIV \ and \HeII\ relative to \MgII, respectively. The 
intrinsic scatter is $374\pm 16\ \mathrm{km\ s^{-1}}$. }
\label{fig:heii_vof}
\end{figure}

\begin{figure}
\epsscale{1.2}
\plotone{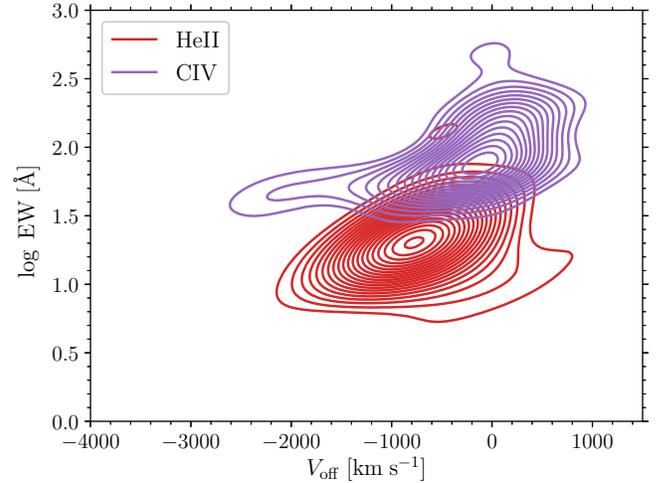}
\caption{Distribution of our sample in the offset velocity-EW plane. \CIV\ and \HeII\ share similarly 
shaped two dimensional distributions. }
\label{fig:heii_cot}
\end{figure}

It is interesting that another high ionization line, \HeII, displays similar 
properties. Indeed, the \HeII\ blueshift (with respect to \MgII) and that of \CIV \ are strongly 
correlated (Figure~\ref{fig:heii_vof}). We adopted the Bayesian linear regression method in 
\cite{Kelly2007} to fit the data. The best-fit relation is $y=(0.69\pm 0.02)x-(583\pm 20)\ 
\mathrm{km\ s^{-1}}$ with an intrinsic scatter of $374\pm 16\ \mathrm{km\ s^{-1}}$, where $x$ 
and $y$ correspond to the blueshifts of \CIV \ and \HeII\ relative to \MgII, respectively.\footnote{Unlike 
\cite{Denney2016a} who tried only to include the narrow component, 
we considered the full \HeII\ line (i.e., the summation of broad and narrow profiles). Therefore, 
it is not straightforward to compare our results with those of \cite{Denney2016a}.} The blueshift 
velocity of \HeII\ is statistically larger than that of \CIV.\footnote{However, this conclusion depends 
on the definition of the shift velocity. If we measure the shift velocity via the line peak and do not 
exclude the components with the ratio of their flux to the total flux $<0.05$, the blueshift 
velocity of \HeII\ is statistically similar to that of \CIV\ \citep{Shen2016}.} Meanwhile, the slope 
of the correlation is shallower than the one-to-one relation. Hence, for sources with \CIV\ $V_{\mathrm{off}} 
<-1000\ \mathrm{km\ s^{-1}}$, the blueshift velocities of \CIV\ are larger than those of \HeII. 
The distribution of our sample in the \HeII\ EW-offset parameter space is similar to that of \CIV \ 
(Figure~\ref{fig:heii_cot}). Our correlation is unlikely to be due to possible redshift biases because, 
similar to those of \CIV, sources with extreme \HeII\ blueshifts also tend to have small EWs. Instead, 
these similarities suggest that the blueshifts of \HeII\ and \CIV \ share the same physical origin, and 
the blueshift might be a common feature of all high-ionization emission lines. 

\begin{figure*}
\epsscale{0.8}
\plotone{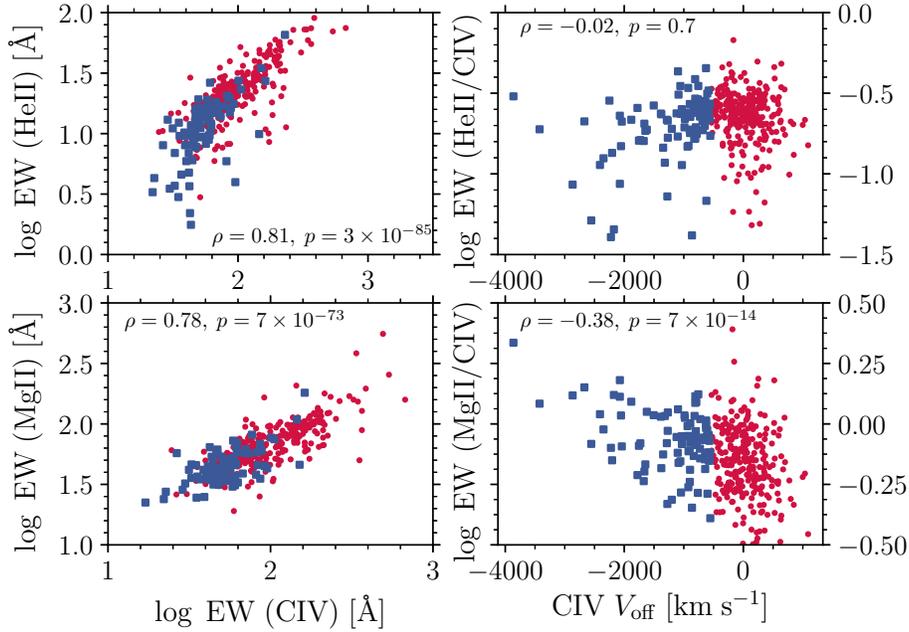}
\caption{Upper-left: \HeII\ EW as a function of \CIV \ EW. Upper-right: $\log \mathrm{EW(HeII)} 
- \log \mathrm{EW(CIV)}$ as a function of the \CIV \ blueshift. Lower panels: the same as the 
upper panels, but for \MgII. In each panel, the high-blueshift sources are highlighted 
as blue squares, and the correlation between the $x$-axis and $y$-axis variables is 
evaluated via the Spearman rank correlation (i.e., Spearman's $\rho$). It appears 
that, in presence of a large \CIV \ blueshift, \CIV \ and \HeII\ are preferentially suppressed with respect 
to \MgII . }
\label{fig:ew_comp}
\end{figure*}

\CIV \ EW correlates well with both \MgII \ EW and \HeII\ EW (left panels of Figure~\ref{fig:ew_comp}). 
As a result, both  \MgII \ EW and \HeII\ EW are correlated with the \CIV\ blueshift. 
Meanwhile, the difference between \CIV \ and \MgII \ EWs and the \CIV \ blueshift are 
anti-correlated (the lower right panel of Figure~\ref{fig:ew_comp}). Such anti-correlation is also obtained 
between \Hbeta\ and \CIV\ \citep{Sulentic2017}. A similar anti-correlation is not evident 
for \CIV \ and \HeII\ (the upper right panel of Figure~\ref{fig:ew_comp}). Compared with \MgII\ (whose 
ionization energy is $\sim 10\ \mathrm{eV}$), the ionization energy of high-ionization BELs (e.g., \CIV\ 
or \HeII) is much higher (i.e., $\sim 50\ \mathrm{eV}$). The ratio of the flux of a BEL to that of quasar 
continuum at the ionization energy measures the covering factor of the BLR gas. If the ratio of the covering 
factor of \MgII\ to that of \CIV\ (or \HeII) is independent of the \CIV\ blueshift, our results indicate that, 
with the presence of a large \CIV \ blueshift, the high-energy ($E\sim 50\ \rm{eV}$) ionizing continuum 
is preferentially reduced with respect to the lower-energy ($E\sim 10\ \rm{eV}$) one. 

\subsection{The blueshift and line widths}
\label{sect:width}
\begin{figure}
\epsscale{1.2}
\plotone{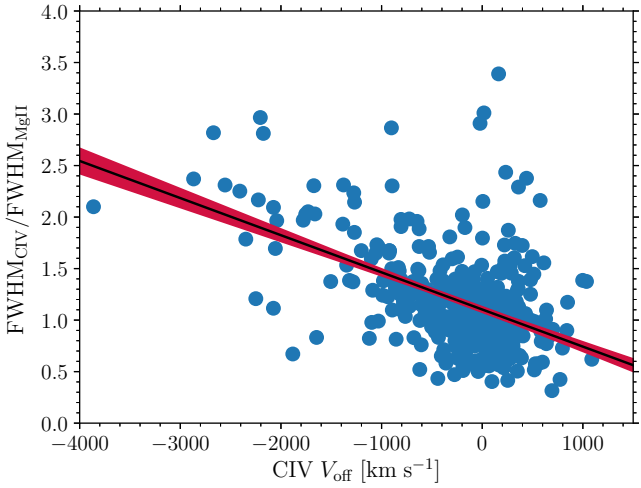}
\caption{Following \cite{Coatman2017}, we fit $\rm{FWHM}_{\rm{CIV}}/\rm{FWHM}_{\rm{MgII}}$ 
as a function of the \CIV \ blueshifts. The best-fit relation \citep[via the Bayesian linear regression method; 
see][]{Kelly2007} is $y = (-0.36\pm 0.03)x + (1.1\pm 0.02)$ with an intrinsic scatter of $0.41\pm 0.02$, where 
$x=V_{\rm{off}}/10^3\ \mathrm{km\ s^{-1}}$. }
\label{fig:fwhm_vof}
\end{figure}

\begin{figure}
\epsscale{1.2}
\plotone{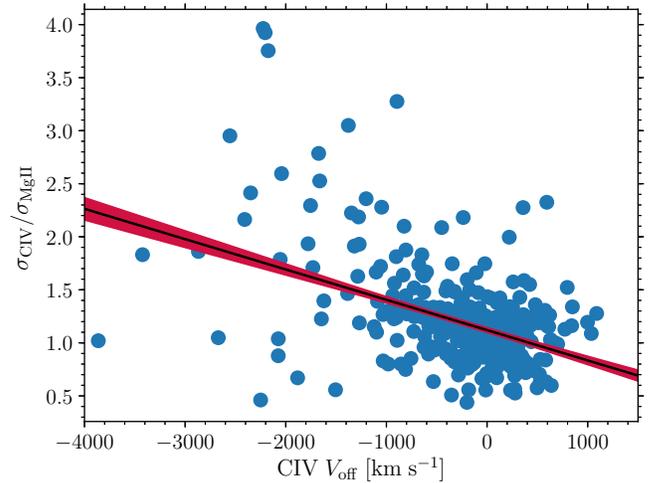}
\caption{The behavior of $\sigma_{\rm{CIV}}/\sigma_{\rm{MgII}}$ as a function of the \CIV\ blueshifts. 
The best-fit relation is $y = (-0.28\pm 0.03)x + (1.12\pm 0.02)$ with an intrinsic scatter of $0.34 
\pm 0.02$. For the extreme blueshift sources, there is no clear 
correlation between $\sigma_{\rm{CIV}}/\sigma_{\rm{MgII}}$ and the \CIV \ blueshift. This result, 
along with Figure~\ref{fig:fwhm_vof}, suggests that, for wind dominated sources, the core of the 
\CIV \ profile is preferentially ``broadened''. }
\label{fig:sigma_vof}
\end{figure}

Previous works \citep[e.g.,][]{Sulentic2007, Shen2012, Runnoe2013, Coatman2016, Coatman2017} 
often argue that, unlike low-ionization emission lines (e.g., \Hbeta, \MgII), \CIV \ is a biased estimator 
of $M_{\rm{BH}}$. Notably, there is an anti-correlation between $\mathrm{FWHM_{CIV}}/\mathrm{FWHM_{MgII}}$ 
and the \CIV \ blueshift \citep[see, e.g.,][]{Shen2008}. We also used our data to explore this 
anti-correlation since our coadded data do not suffer from short-timescale (i.e., rest-frame 
$\sim 60$ days) quasar variability. We confirm an anti-correlation between 
$\mathrm{FWHM_{CIV}}/\mathrm{FWHM_{MgII}}$ and the \CIV \ 
blueshift (Figure~\ref{fig:fwhm_vof}), indicating that MgII- and CIV-based $M_{\rm{BH}}$ will be inconsistent. 
Similar relations are also observed between the ratio of the line dispersion 
$\sigma_{\mathrm{CIV}}/\sigma_{\mathrm{MgII}}$ and the \CIV \ blueshift (Figure~\ref{fig:sigma_vof}). 

These results suggest that, in the presence of the \CIV \ blueshift, both $\mathrm{FWHM_{CIV}}$ and 
$\sigma_{\mathrm{CIV}}$ are biased, and corrections are required. Unlike FWHM, the line dispersion 
$\sigma$ is more sensitive to the wings of the line profile. The correction for 
$\sigma_{\mathrm{CIV}}$ is much smaller than that for $\mathrm{FWHM_{CIV}}$. Therefore, the core of the 
\CIV \ profile is preferentially ``broadened'' as a function of the \CIV \ blueshift. This result 
also indicates that $\sigma$ is a more reliable estimator of the virial motions. In practice, 
we prefer to estimate $M_{\rm BH}$ by making empirical corrections to $\mathrm{FWHM_{CIV}}$ \citep[see 
also][]{Coatman2017}. The proposed approach has two advantages. First, FWHM is better 
constrained than $\sigma$ in low S/N spectra. Second, for extreme blueshift sources, there is no clear 
correlation between $\sigma$ and the blueshift. 

The slope of our anti-correlation between the ratio of $\mathrm{FWHM_{CIV}}$ to $\mathrm{FWHM_{MgII}}$ 
and the \CIV \ blueshift is consistent with the correlation between \CIV \ and \Hbeta\ found by 
\cite{Coatman2017}. The intercept and the intrinsic scatter of our relation are larger than those of 
\cite{Coatman2017} by a factor of two. The differences could be caused by the imperfect one-to-one relation 
between $\mathrm{FWHM_{MgII}}$ and $\mathrm{FWHM_{H\beta}}$ \citep{Trakhtenbrot2012} or the differences 
in how we estimate \CIV\ blueshift compared with \cite{Coatman2017}. 

\begin{figure}
\epsscale{1.2}
\plotone{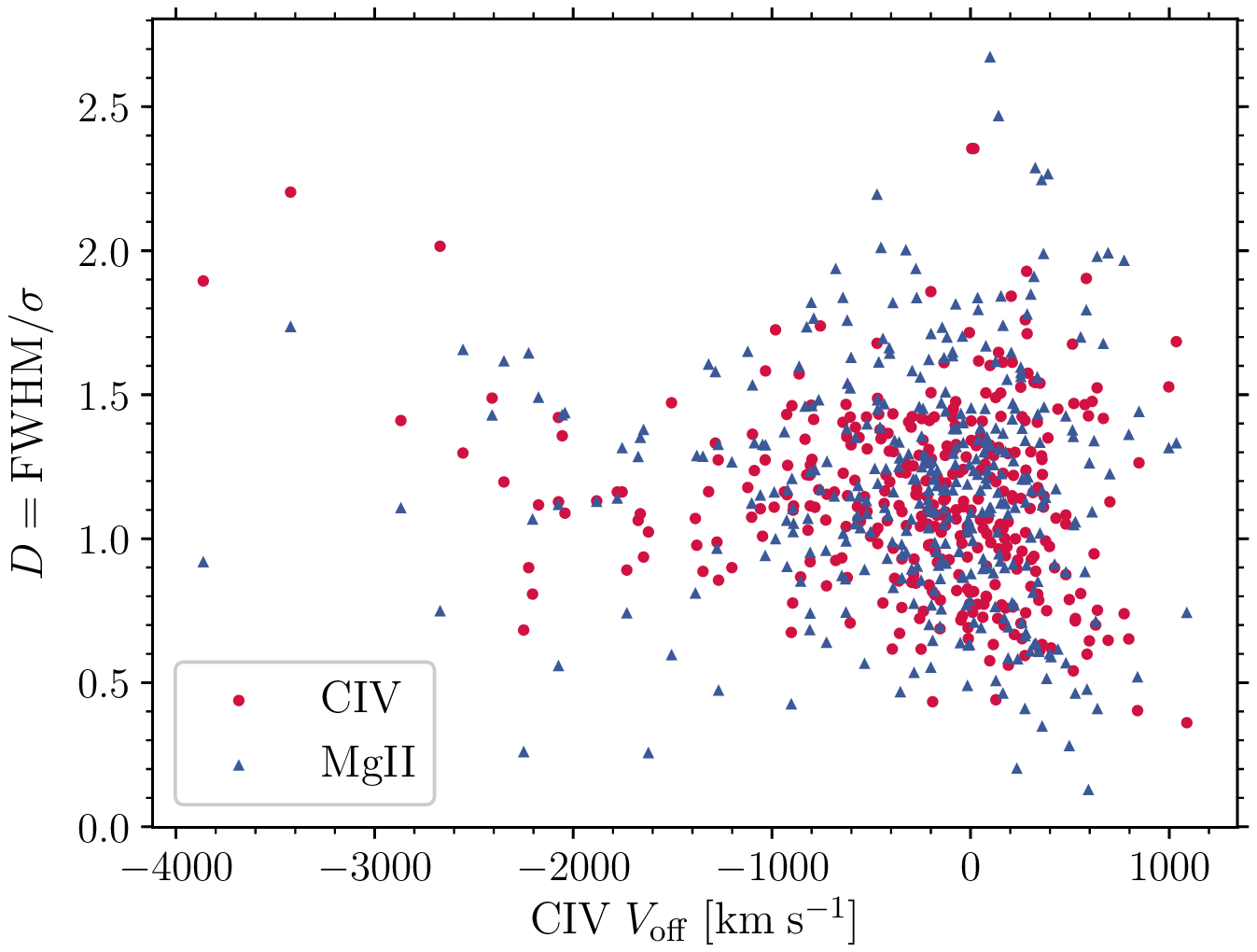}
\caption{The behavior of the line shape $D$ as a function of the \CIV\ blueshifts. For \CIV, there is a 
weak anti-correlation between $D_{\mathrm{CIV}}$ and the \CIV\ blueshifts ($\rho=-0.13$ and $p=0.008$). 
However, there is no anti-correlation between $D_{\mathrm{MgII}}$ and the \CIV\ blueshifts ($\rho=-0.07$ 
and $p=0.1$).}
\label{fig:plf_vof}
\end{figure}

The shape of \CIV, $D_{\mathrm{CIV}}$ (for its definition, see Section~\ref{sect:emlp}), is also 
expected to be anti-correlated with the \CIV \ blueshift since the corrections for FWHM and 
$\sigma$ are different. This speculation is confirmed by the Spearman rank correlation 
($\rho=-0.13$, and $p=0.008$; see Figure~\ref{fig:plf_vof}), albeit with a substantial 
scatter. In other words, the high-blueshift sources tend to avoid the small $D_{\mathrm{CIV}}$ 
(i.e., more boxy) space. Therefore, like the \CIV \ blueshift, $D_{\mathrm{CIV}}$ 
can also be adopted as a viable and practical proxy to correct $\mathrm{FWHM_{CIV}}$ \citep{Denney2012}, 
although the correction may be inadequate at extreme \CIV \ blueshift. There are two possible 
explanations for the anti-correlation. First, the spectra of the low-blueshift sources have significant 
narrow \CIV\ components. The contribution of the narrow \CIV\ component to the total flux is smaller 
for the high-blueshift sources. If this scenario is correct, we would expect the shape of \MgII\ to show 
the same anti-correlation. However, we found that, according to the Spearman rank correlation 
test, there is no significant correlation between the shape of \MgII\ and the \CIV\ blueshift ($\rho=-0.07$ 
and $p=0.1$; see Figure~\ref{fig:plf_vof}). Second, there is an intrinsic anti-correlation between the 
shape of \CIV\ and its blueshift. This anti-correlation seems to be inconsistent with the scenario 
proposed by \cite{Gaskell2009} where the line profiles of high-blueshift sources are expected 
to be less boxy (i.e., have smaller values of $D_{\mathrm{CIV}}$). 

\section{Variability of the line shift}
\label{sect:result_sp}
We can also study the \CIV \ blueshift for each of the $29$ epochs. The intrinsic variability of the 
blueshift of \CIV \ can be constrained by calculating the ``excess of variance'' \citep[see, e.g.,][]{Sun2015}, 
\begin{equation}
\label{eq:iqr}
\mathrm{VAR.}(V_{\mathrm{shift,se}}) = \sqrt{(0.74\mathrm{IQR}(V_{\mathrm{shift,se}}))^2 - 
\widetilde{V^2}_{\mathrm{err}}} \,
\end{equation}
where $\mathrm{IQR}(V_{\mathrm{shift,se}})$ and $V_{\mathrm{err}}$ are the $25\%-75\%$ interquartile 
range and the uncertainty of $V_{\mathrm{shift,se}}$, respectively. $\widetilde{V^2}_{\mathrm{err}}$ 
represents the median value of the variable $V^2_{\mathrm{err}}$. 

The dynamical timescale of the BLR is 
\begin{equation}
T_{\mathrm{dyn}} \sim \frac{2\pi}{\Omega_{\mathrm{K}}} = 253(\frac{R_\mathrm{BLR}}{10^3R_{\mathrm{S}}})^{3/2}
\frac{M_{\mathrm{BH}}}{5\times 10^7M_{\odot}}\ \mathrm{days}
\end{equation}
where $\Omega_{\mathrm{K}}$, $R_\mathrm{BLR}$ and $R_\mathrm{S}$ are the Keplerian angular velocity, the 
radial distance of the BLR to the SMBH, and the Schwarzschild radius, respectively. With timescales we consider 
here (i.e., $\lesssim 60$ rest-frame days, which are much smaller than $T_{\mathrm{dyn}}$), the variability is 
likely driven by the quasar continuum variations. For instance, let us consider that the time lags 
between the ionizing continuum and the BELs depend on the line-of-sight velocities \citep[see, e.g.,][]{Denney2009, 
Grier2013}. The time lags of the blue and red wings will be different. If the time lag of the blue wing is shorter 
than the red wing, as the quasar continuum increases (decreases), we will observe a quick response of the blue 
wing, i.e., an apparent blue (red) shift until the red wing responds to the increase (decrease) at a later epoch. 
Therefore, being either blueshifted or redshifted, and the variations on short timescales are expected 
\citep[hereafter the ``line-of-sight velocity-dependent reverberation''; see][]{Barth2015}. 

\subsection{The line shift of \MgII}
\label{sect:varmg}
\begin{figure}
\epsscale{1.2}
\plotone{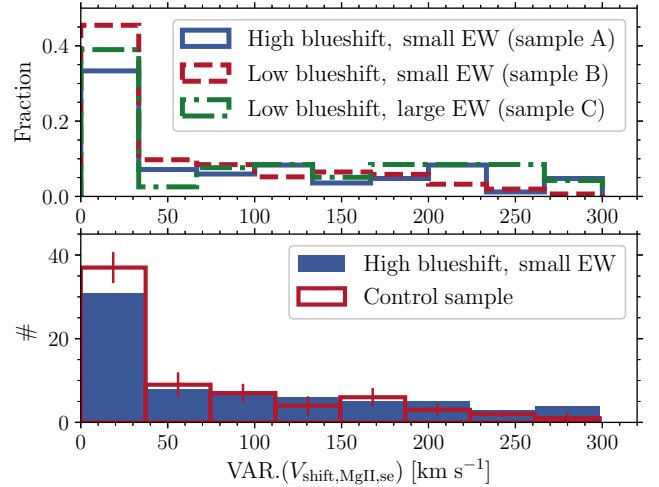}
\caption{Upper panel: Distributions of the intrinsic variability of $V_{\rm{shift,MgII,se}}$ for three different 
subsamples. Most sources show weak variability of $V_{\rm{shift,MgII,se}}$. 
Lower panel: A comparison between the distribution of the \MgII \ line shift variations of sample 
A sources (i.e., the ``high-blueshift, small-EW'' sources) to that of the ``controlled sample'' 
(matched in \CIV \ EW, $L1350$, and redshift). The median VAR.($V_{\rm{shift,MgII,se}}$) of extreme blueshift 
sources is larger than that of the control sample. }
\label{fig:mgii_v50_var}
\end{figure}

To understand the line-shift variations due to reverberation, we first study \MgII. The upper panel 
of Figure~\ref{fig:mgii_v50_var} illustrates the distribution of the variability of the line shift 
of \MgII \ (VAR.($V_{\rm{shift, MgII, se}}$)) for our three subsamples. Note that the uncertainty of 
VAR.($V_{\rm{shift, MgII, se}}$) is generally not small for each source. Hence, we focus only on the 
median VAR.($V_{\rm{shift, MgII, se}}$) of each subsample. Meanwhile, we have visually inspected 
those sources that show very large VAR.($V_{\rm{shift, MgII, se}}$) (i.e., $\gg 300\ \mathrm{km\ s^{-1}}$) 
and found that, in many cases, their single-epoch spectra are noisy and the best-fitting results are not 
robust. Therefore, we rejected these sources. 

To better control the effect of EW(\CIV), we created control samples matched in EW(\CIV), $L1350$, 
and redshift. More specifically, for each source in Sample A (i.e., the ``high-blueshift, small-EW" 
sample), we randomly (with replacement) selected a quasar from the sources in Samples B or C with 
similar \CIV \ EW (within $0.1$ dex). We then adopt the Anderson-Darling test\footnote{The Anderson-Darling 
test is found to be more sensitive at recognizing the difference between two distributions than the 
popular Kolmogorov-Smirnov test \citep[e.g.,][]{Hou2009}.} to measure the probability that the randomly-selected 
sample is drawn from the same parent population as sample A in terms of \CIV \ EW, $L1350$, and 
redshift. The two samples are consistent with being matched if the null-hypothesis probability $p>0.05$.
We then calculate the distributions of VAR.($V_{\rm{shift, MgII, se}}$) for sample A and the control 
sample. We repeated this procedure $1024$ times. Our results are presented in the lower 
panel of Figure~\ref{fig:mgii_v50_var}. The median VAR.($V_{\rm{shift, MgII, se}}$) 
of sample A is $56.4\pm 8.9\ \mathrm{km\ s^{-1}}$, whereas, the median VAR.($V_{\rm{shift, MgII, se}}$) 
of the control sample is $33.1\pm 16.5\ \mathrm{km\ s^{-1}}$.\footnote{The uncertainty is the dispersion 
of the $1024$ realizations.} Of our $1024$ realizations, the possibility to have VAR.($V_{\rm{shift, MgII, se}}$) $\geq 
56.4\ \mathrm{km\ s^{-1}}$ is less than $3\%$. We thus conclude that VAR.($V_{\rm{shift, MgII, se}}$) of sample 
A is statistically larger than that of the control sample. 

\subsection{The line shift of \CIV}
\label{sect:varciv}
In this section, we check the variability properties of the \CIV\ blueshift with respect to the coadded 
high S/N \MgII. Therefore, the variations presented in this section are only due to the line shift of \CIV. 
As in Section~\ref{sect:varmg}, we focus only on the median VAR.($V_{\rm{shift, CIV, se}}$) , discarding 
sources with VAR.($V_{\rm{shift, CIV, se}}$) $>300\ \mathrm{km\ s^{-1}}$. 

\begin{figure}
\epsscale{1.2}
\plotone{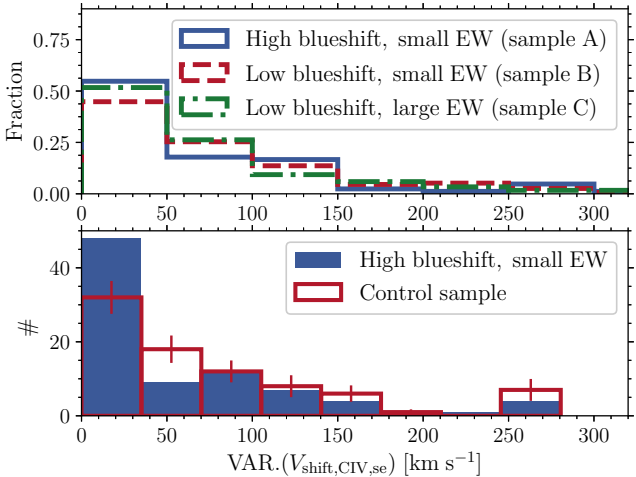}
\caption{Upper panel: Distributions of the intrinsic variability of $V_{\rm{shift,CIV,se}}$ for three different subsamples. 
Similar to that of \MgII , most sources show weak variability of $V_{\rm{shift,CIV,se}}$.
Lower panel: A comparison between the distribution of the \CIV \ line shift variations of sample A 
sources (i.e., the ``high blueshift, small EW'' sources) and that of ``controlled sample'' (matched in 
\CIV \ EW, $L1350$, and redshift). The high-blueshift sources have smaller median VAR.($V_{\rm{shift,CIV,se}}$), 
opposite to what is observed for \MgII.}
\label{fig:civ_v50_var}
\end{figure}

Figure~\ref{fig:civ_v50_var} shows the results of the analysis of VAR.($V_{\rm{shift,CIV,se}}$). 
The median VAR.($V_{\rm{shift,CIV,se}}$) of Sample A (i.e, the ``high blueshift, small EW" sample) 
is $11.6\pm 4.5\ \mathrm{km\ s^{-1}}$. For the control sample, the median VAR.($V_{\rm{shift,CIV,se}}$) 
is $58.7\pm 8.7\ \mathrm{km\ s^{-1}}$. Of our $1024$ realizations, none has VAR.($V_{\rm{shift,CIV,se}}$) 
$\leq 11.6\ \mathrm{km\ s^{-1}}$. Therefore, contrary to the results for \MgII, sample 
A sources have slightly weaker \CIV \ line shift variability than that of the control sample. Meanwhile, 
the line shift of \MgII \ varies more strongly than that of \CIV. Indeed, the Mann-Whitney U test\footnote{The 
Mann-Whitney test is a nonparametric test of the null hypothesis that the distributions of two populations are 
equal.} of the line-shift variability on \MgII\ and \CIV\ indicates that the variation amplitude of \MgII\ is 
statistically larger than that of \CIV\ (the $p$-value is $0.003$). These differences indicate that the structure 
of \CIV\ evolves as a function of the \CIV\ blueshift in a way that is different from that of \MgII.

In the high-redshift universe, broad \MgII \ or \CIV \ lines are sometimes adopted to determine 
the quasar redshift. As discussed by \cite{Denney2016a, Denney2016b} and \cite{Shen2016}, such 
redshift estimation is significantly biased, depending on quasar properties. The bias can be 
corrected to be better than $\sim 200\ \rm{km\ s^{-1}}$ by some empirical guidelines \citep{Shen2016}. 
However, our results (Figures~\ref{fig:mgii_v50_var} and ~\ref{fig:civ_v50_var}) indicate that 
quasar variability places lower limits on the accuracy of measuring quasar redshifts 
with only single-epoch BELs. For some sources, the accuracy can be worse than $200\ 
\rm{km\ s^{-1}}$.

\section{Discussion}
\label{sect:dis}
\subsection{\CIV \ blueshift and quasar variability}
\label{sect:dis1}
Our analyses in Section~\ref{sect:result_sp} demonstrate that the observed \CIV \ blueshift can vary due to 
line-of-sight velocity-dependent reverberation. Therefore, quasar variability might have effects on the \CIV \ 
EW-offset velocity connection. 

\begin{figure}
\epsscale{1.2}
\plotone{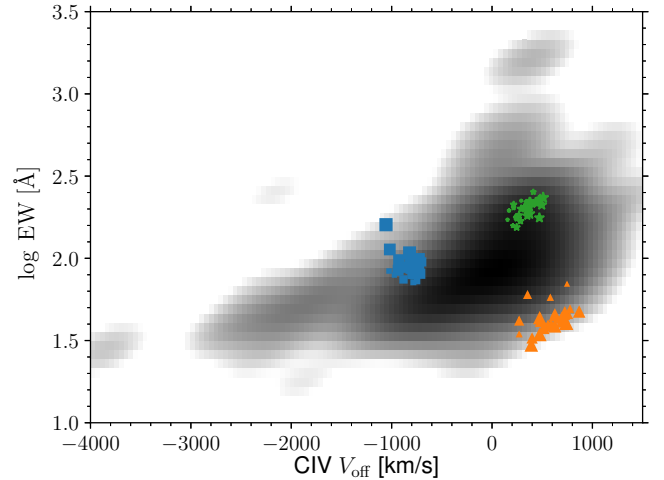}
\caption{The locations of three sources in the \CIV \ EW-offset velocity plane. The three 
sources are selected from sample A (blue squares), sample B (orange triangles) and 
sample C (green stars), respectively. For each source, the symbol size increases with 
time. The grey color indicates the probability density distribution of our whole sample. 
Consistent with Figure~\ref{fig:civ_v50_var}, sources do not strongly change their 
positions. }
\label{fig:particle}
\end{figure}

Let us first determine how much quasars can move in the \CIV \ EW-offset velocity plane due to quasar 
variability. To do so, we selected three sources, i.e., one from sample A, one from sample B, and one from 
sample C. The selection criteria are as follows. For each sample, we selected the source with 
the highest ratio of the VAR.($V_{\rm{shift,CIV,se}}$) to the median measurement error of \CIV \ 
$V_{\rm{shift,CIV,se}}$ such that VAR.($V_{\rm{shift,CIV,se}}$) is the most robust one. 
Figure~\ref{fig:particle} presents our results. For the three selected sources, $\sim 20\%$ of their motions 
along the \CIV\ blueshift are due to measurement uncertainties. These sources do not rapidly 
change their positions over the time period of our observations (see also Figure~\ref{fig:civ_v50_var}). 

\begin{figure}
\epsscale{1.2}
\plotone{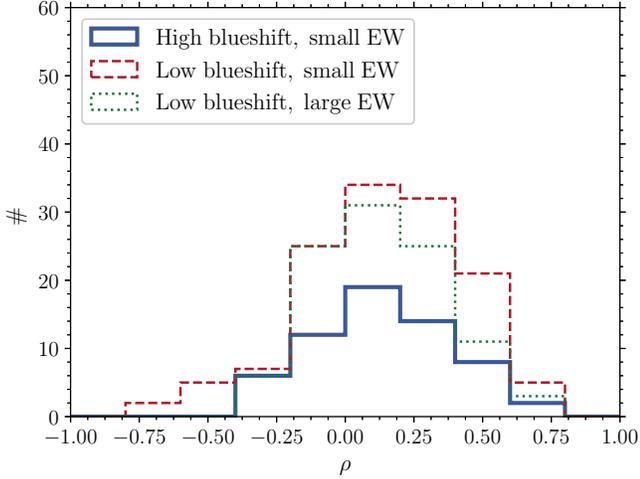}
\caption{The distributions of the Spearman rank correlation coefficient $\rho$ between 
$V_{\mathrm{off,se}}$ and EW over the $29$ epochs. There is, on average, a positive 
correlation between $V_{\mathrm{off,se}}$ and EW, which is consistent with the global 
\CIV \ EW-offset connection. }
\label{fig:par_cor}
\end{figure}

\begin{figure}
\epsscale{1.2}
\plotone{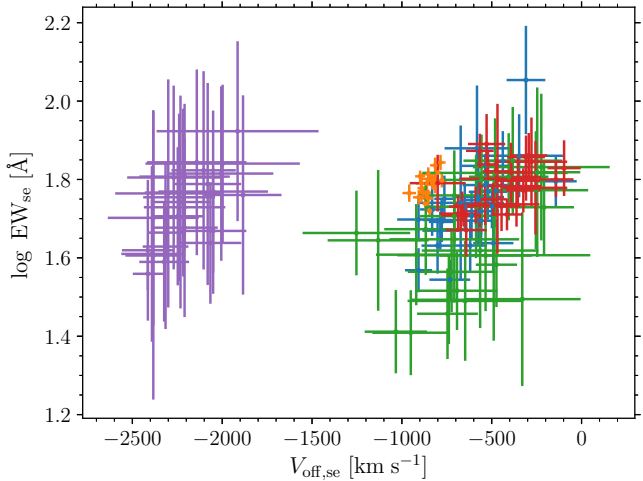}
\caption{The single-epoch \CIV \ EW as a function of $V_{\rm{off,se}}$ for the five sources 
in sample A (i.e., the ``high-blueshift, small-EW'' sample). For these sources, there are statistically 
significant positive correlations between the single-epoch \CIV \ EW and $V_{\rm{off,se}}$. }
\label{fig:lcs_A}
\end{figure}

\begin{figure}
\epsscale{1.2}
\plotone{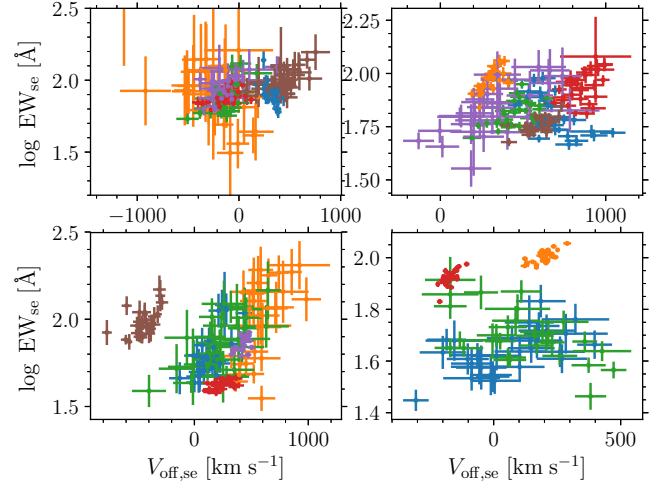}
\caption{The single-epoch \CIV \ EW as a function of $V_{\rm{off,se}}$ for the $22$ sources 
in sample B (the ``low-blueshift, small-EW'' sample). For these sources, there are statistically 
significant positive (eighteen sources) or negative (four sources) correlations between the 
single-epoch \CIV \ EW and $V_{\rm{off,se}}$. To avoid 
severe overlapping and confusion, the $22$ sources spread across four panels. Each of 
the upper and lower-left panels contains six sources; the lower-right panel contains four 
sources. }
\label{fig:lcs_B}
\end{figure}

\begin{figure}
\epsscale{1.2}
\plotone{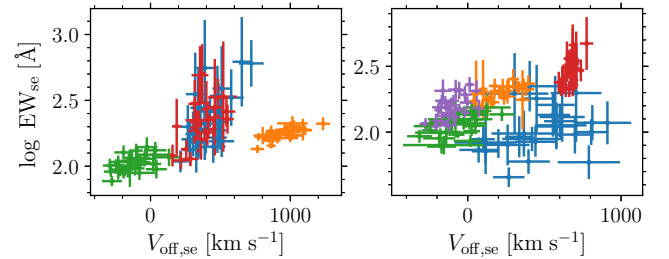}
\caption{The single-epoch \CIV \ EW as a function of $V_{\rm{off,se}}$ for the nine sources 
in sample C (the ``low-blueshift, large-EW'' sample). For these sources, there are statistically 
significant positive correlations between the single-epoch \CIV \ EW and $V_{\rm{off,se}}$. To 
avoid severe overlapping and confusion, the nine sources spread across two panels. The left 
(right) panel contains five (four) sources. }
\label{fig:lcs_C}
\end{figure}

We then explored the possible correlation between $V_{\rm{off,se}}$ and EW over the $29$ epochs 
for each source. Figure~\ref{fig:par_cor} presents the distributions of the Spearman rank correlation 
coefficient $\rho$. For most of our sources, the correlation is statistically insignificant. This result is 
not totally unexpected as the timescale of our multi-epoch data is short ($\sim 60$ days) and quasar 
variability is weak on short timescales. However, on average, 
$V_{\rm{off,se}}$ and EW are positively correlated over the $29$ epochs. 
We examined further the $36$ sources with statistically significant correlations and found that only 
four out of them have negative correlations ($0/5$, $4/22$, and $0/9$ for samples A, B, and C, 
respectively). In Figures~\ref{fig:lcs_A}, ~\ref{fig:lcs_B}, and ~\ref{fig:lcs_C}, we illustrate the 
single-epoch \CIV \ EW as a function of $V_{\rm{off,se}}$ for these $36$ sources. Most of their 
motions along the \CIV\ EW are due to random fluctuations that are driven by the measurement 
uncertainties. 

\begin{figure}
\epsscale{1.2}
\plotone{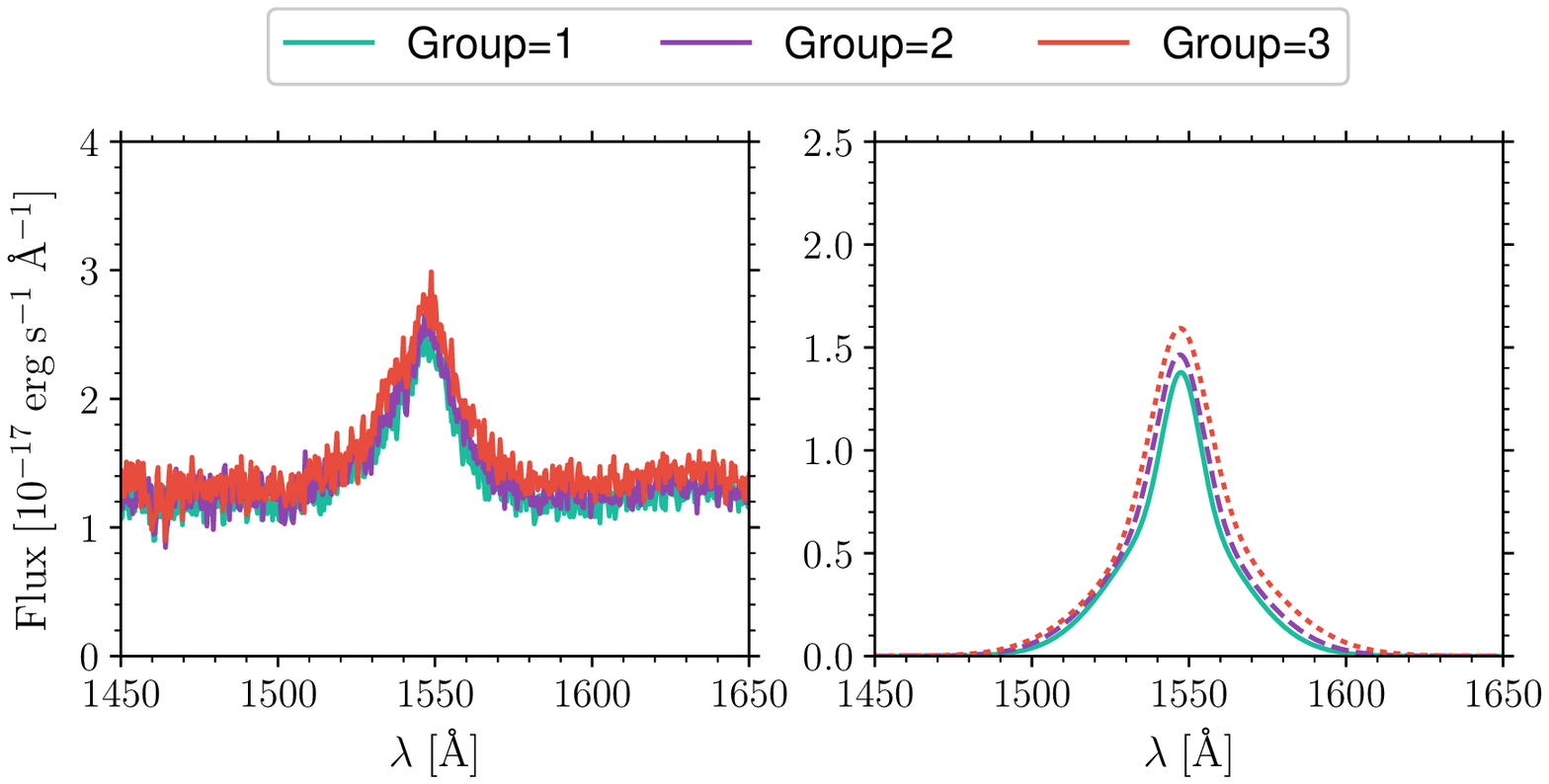}
\caption{The time evolution of the \CIV\ profile for a sample A QSO (RMID=$108$). The left and 
right panels represent the high S/N mean spectrum in each group (for its definition, see texts for 
more details) and the best fits of \CIV . The spectra and the best fits of \CIV\ profiles are normalized 
to the best-fitting $1550\ \mathrm{\AA}$ continuum. In this example, groups with higher 
\CIV\ EW tend to be less blueshifted.}
\label{fig:RM_A}
\end{figure}

\begin{figure}
\epsscale{1.2}
\plotone{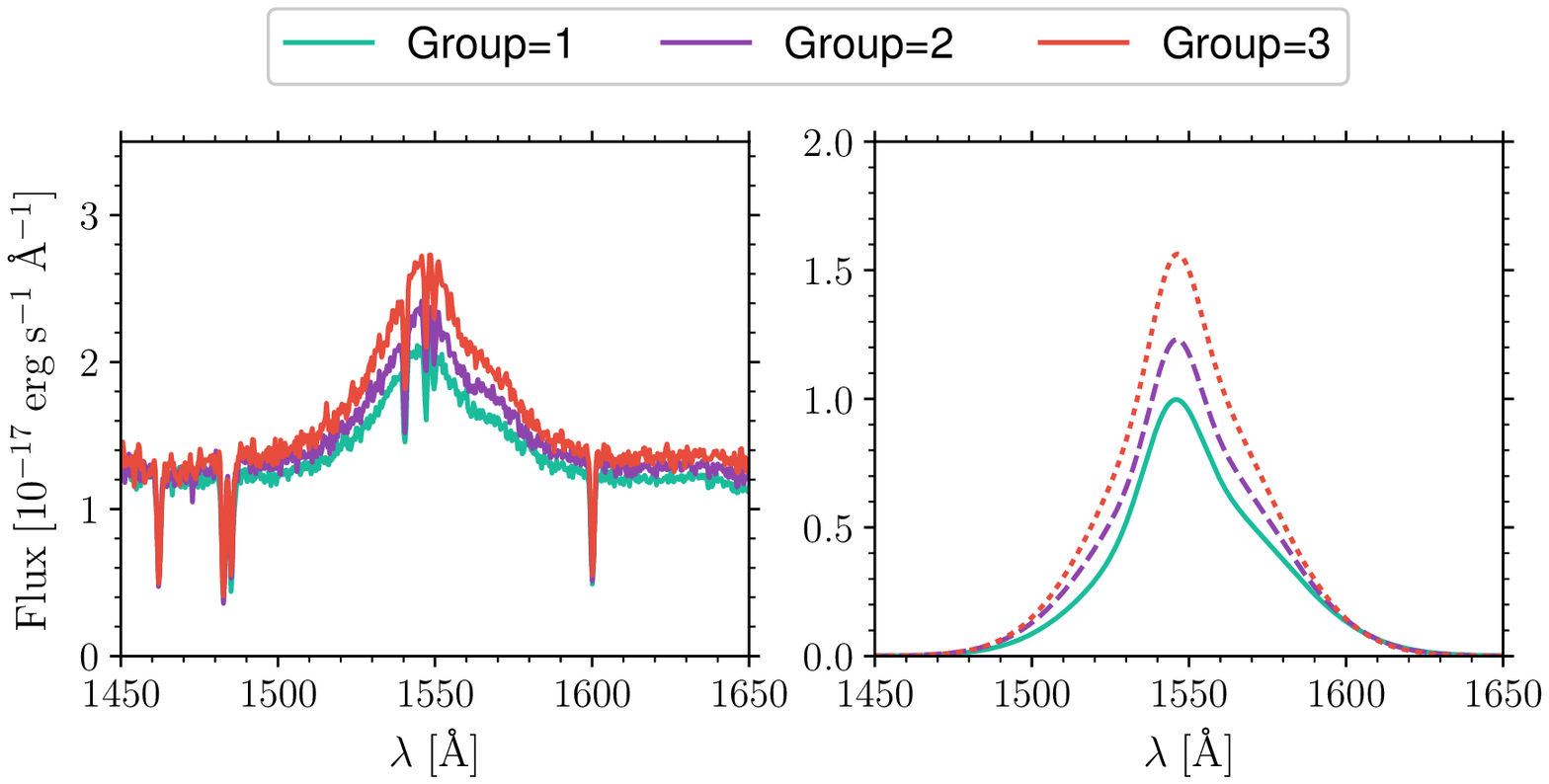}
\caption{The time evolution of the \CIV\ profile for a sample B QSO (RMID=$231$). The left and 
right panels represent the high S/N mean spectrum in each group (for its definition, see texts for 
more details) and the best fits of \CIV . The spectra and the best fits of \CIV\ profiles are normalized 
to the best-fitting $1550\ \mathrm{\AA}$ continuum. In this example, groups with higher 
\CIV\ EW tend to be more blueshifted.}
\label{fig:RM_B}
\end{figure}

\begin{figure}
\epsscale{1.2}
\plotone{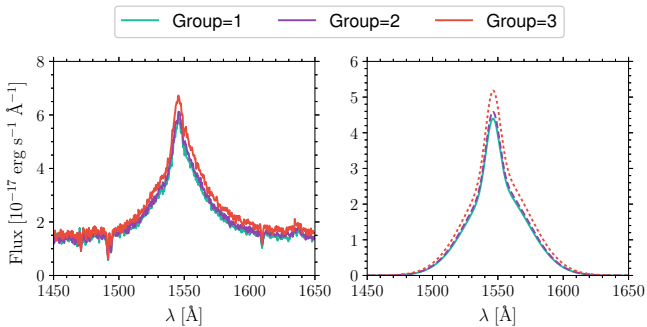}
\caption{The time evolution of the \CIV\ profile for a sample C QSO (RMID=$491$). The left and 
right panels represent the high S/N mean spectrum in each group (for its definition, see texts for 
more details) and the best fits of \CIV . The spectra and the best fits of \CIV\ profiles are normalized 
to the best-fitting $1550\ \mathrm{\AA}$ continuum. In this example, groups with higher 
\CIV\ EW tend to be less blueshifted.}
\label{fig:RM_C}
\end{figure}

In Figures~\ref{fig:RM_A}, ~\ref{fig:RM_B}, and ~\ref{fig:RM_C}, we show three examples of the 
variations of the \CIV\ profile. As an aid to visual inspection, for each source, we divided its 
$29$ observations into three groups according to the increasing single-epoch \CIV 
\ EW (i.e., the $29$ observations are sorted by the single-epoch \CIV \ EW; 
the $1$st-$9$th, $10$th-$19$th, and $20$th-$29$th re-ordered observations 
belong to Groups $1$, $2$, and $3$, respectively); for each group, we created the variance-weighted 
high S/N mean spectrum; we fitted the mean spectra following the spectra-fitting approach mentioned 
in Section~\ref{sect:spfit}. When being plotted, all spectra and best fits of \CIV\ profiles are 
normalized to the best-fitting $1550\ \mathrm{\AA}$ continuum. Therefore, the intensities of \CIV\ 
in the figures are proportional to their EWs. 

The positive correlation between \CIV\ EW and offset velocity could be induced by some bias in our 
spectral fitting procedure. As mentioned in Section~\ref{sect:linefit}, we only fit \CIV\ in [$1500\ 
\mathrm{\AA}$, $1600\ \mathrm{\AA}$] for single epoch spectra. It is possible that out code interprets 
weak \HeII\ as the red wing of \CIV. As a result, we would expect 
an artificial positive correlation between \CIV\ EW and offset velocity. We therefore performed a simple 
simulation to account for this bias. We selected RMID=$693$ as an example.\footnote{We chose this 
source because the boundary between \CIV\ and \HeII\ is not visually evident in its composite spectrum. 
Therefore, the bias can be large. We also tested some other sources and found similar results. } A total 
of $29$ mock spectra were generated, where the flux in each 
wavelength pixel was determined by adding the single-epoch flux density noise to the best-fit model of 
the composite spectrum. We then fitted the mock spectra following the same fitting recipe. We find that 
the variability in the mock spectra is mostly due to measurement errors and the variability amplitude 
is much less than the true single-epoch spectra. In addition, for the mock spectra, the correlation between 
\CIV\ EW and offset velocity is statistically insignificant. We therefore conclude that the bias we mentioned 
cannot be responsible for the observed positive correlation. 

How do we understand these statistically significant positive or negative correlations? 
As we mentioned in Section~\ref{sect:result_sp}, the observed line-shift variations are likely driven 
by the ``line-of-sight velocity-dependent reverberation''. Meanwhile, as revealed by many multi-wavelength 
variability studies, quasars tend to be bluer when they become brighter \citep[e.g.,][]{Giveon1999, Berk2004, 
Guo2016} and such a behavior is more prominent on short timescales \citep[$\sim 10$ days, see][]{Sun2014, 
Zhu2016}. Let us again image that the time lag of the blue part is shorter than that of the red one. As 
the quasar continuum increases (decreases), the blue wing will respond faster than the red wing, which results 
in an apparent blueshift (redshift); we will observe a harder (softer) quasar SED, i.e., a larger (smaller) 
EW. Therefore, these positive or negative correlations might be driven by the dependency of the BEL time 
lag on the line-of-sight velocity and the color variability of quasars. 

\begin{figure}
\epsscale{1.2}
\plotone{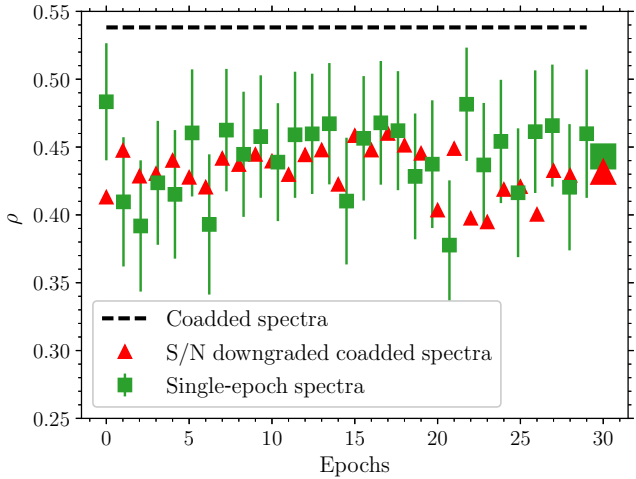}
\caption{The Spearman rank correlation coefficient, $\rho$, between the \CIV \ EW and offset velocity 
for each epoch (small green squares). The black dashed line corresponds to the measurements of the high 
S/N composite spectra. Small red triangles are for the measurements of the composite spectra with 
downgraded S/N (i.e., matched in the S/N of each single-epoch measurements). The large red triangle and 
the large green square represent the mean correlation coefficient for the S/N downgraded composite data 
and the single-epoch data, respectively. The green squares are, on average, above the red triangles, which 
indicates that the correlation is slightly tighter for the single-epoch data. That is, quasar 
variability can enhance the connection between the \CIV \ EW and offset velocity. }
\label{fig:var_eff}
\end{figure}

The overall positive correlation is consistent with the global \CIV \ EW-offset velocity connection 
(i.e., sources with strong \CIV\ tend to be less blueshifted; see, e.g., 
Figure~\ref{fig:ew_vof}). As a result, we expect that 
quasar variability acts in such a way to enhance the global \CIV \ EW-offset velocity connection. 
Compared with the single-epoch data, our high S/N composite spectra do not suffer from 
short-timescale ($\sim 60$ rest-frame days) variability. Therefore, we can assess the effect of quasar 
variability by comparing the scatter of the \CIV \ EW-offset velocity connection of the composite 
spectra with that of the single-epoch data. We first added random noise to the \CIV \ EW 
and $V_{\rm{off}}$ (i.e., measurements from the composite spectra) such that their S/N are identical 
to those of the \CIV \ $\mathrm{EW_{se}}$ and $V_{\mathrm{off, se}}$ (i.e., measurements 
from the single-epoch spectra). The \CIV \ velocity offset for each epoch we adopted here 
is with respect to the composite \MgII \ profile (i.e., $V_{\rm{off,se}}$). We adopted this definition to 
focus on the variability of \CIV \ alone. We then calculated Spearman's $\rho$ 
between the \CIV \ EW-offset velocity for the S/N downgraded composite data; Figure~\ref{fig:var_eff} 
presents our results. The exact value of $\rho$ depends on epochs since the S/N of spectra 
changes with epochs. We also obtained the median $\rho$ over the $29$ epochs. The differences between 
the correlation coefficient of the S/N downgraded composite data and that of the single-epoch data are 
due to quasar variability. The median $\rho$ of the single-epoch spectra is slightly larger 
(by $2.5\%$) than that of the S/N downgraded composite data. Our results indicate that 
quasar variability might slightly enhance the connection between the \CIV \ EW-offset velocity. 

\subsection{The physical origin of the \CIV \ blueshift}
\label{sect:dis2}
\begin{figure}
\epsscale{1.2}
\plotone{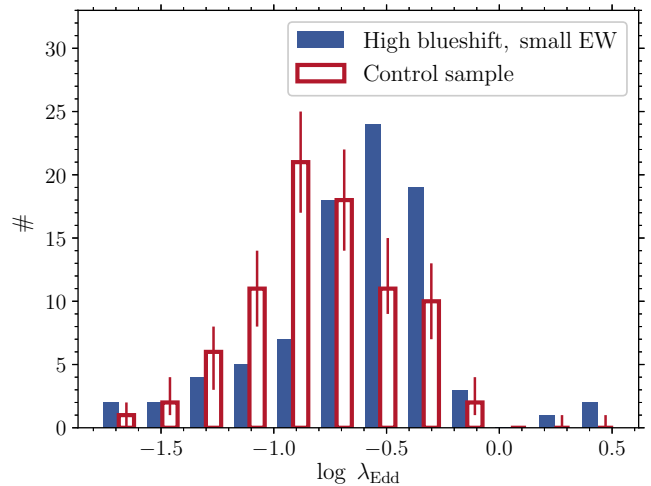}
\caption{The distributions of the Eddington ratio, $\lambda_{\rm{Edd}}$. Filled blue bars represent 
sample A sources (the ``high-blueshift, small-EW'' sources). Open red bars with errors represent the 
control sample (matched in \CIV \ EW, $L1350$, and redshift). The extreme blueshift sources tend to have 
large Eddington ratios. }
\label{fig:ledd_v50}
\end{figure}

It remains unclear why sources with blueshift tend to have low EW, but many other low EW quasars do not 
have a blueshift. We compare these two types of sources in terms of other quasar properties, especially the 
Eddington ratio, as there are suggestions that the \CIV \ EW is tightly correlated with the Eddington ratio 
\citep[e.g.,][]{Bachev2004, Baskin2004, Shemmer2015}. To reduce effects caused by other factors, for Sample 
A (sources with low EW and high blueshift), we again made control samples matched in the \CIV \ EW, 
$L1350$, and redshift. Our results are presented in Fig~\ref{fig:ledd_v50}. While the median logarithmic 
Eddington ratio of sample A is $-0.6$, the median logarithmic Eddington ratio of the control sample is $-0.86 
\pm0.04$. Therefore, after controlling for the \CIV\ EW, quasar luminosity, and redshift, the \CIV \ blueshift 
sources tend to have significantly larger ($\sim 0.3$ dex) Eddington ratios. 

There are two simple scenarios that can explain this result. First, the observed correlation is simply 
an orientation effect \citep{Denney2012}. In this scenario, the high- and low-blueshift sources might 
intrinsically have similar Eddington ratios. However, the high-blueshift sources are viewed more face-on. 
When being viewed face on, the geometrically thin accretion disk will be more luminous than the edge-on 
case. In addition, the face-on systems suffer less from absorption (due to, e.g., a torus) than the 
edge-on ones. For a polar wind, the line-of-sight blueshift velocity would be higher for the face-on 
case. As a result, the extreme blueshift sources apparently have larger Eddington ratios than the 
low blueshift counterparts. This scenario is, however, challenged by some recent observations. For 
instance, \cite{Runnoe2014} measured orientation for a quasar sample via the radio core dominance 
parameter; they found that there is no correlation between the \CIV\ blueshift and orientation. 
Second, the \CIV \ blueshift sources are intrinsically more active, i.e., 
the high-blueshift sources have larger Eddington ratios than those of the low-blueshift sources. 
These two scenarios can be further tested by exploring the variability of the quasar continuum. 
The variability of the quasar optical/UV continuum is observed to be anti-correlated with the Eddington 
ratio \citep[e.g.,][]{Ai2010, Macleod2010, Kelly2013, Kozlowski2016, Rumbaugh2017}, after controlling for 
quasar luminosity, and redshift. Therefore, according to the first scenario, the high- and low-blueshift 
sources share similar variations of the quasar continuum; however, the variability amplitude of the quasar 
continuum for the high-blueshift sources would be smaller than the low-blueshift ones in the second 
scenario. 

\begin{figure}
\epsscale{1.2}
\plotone{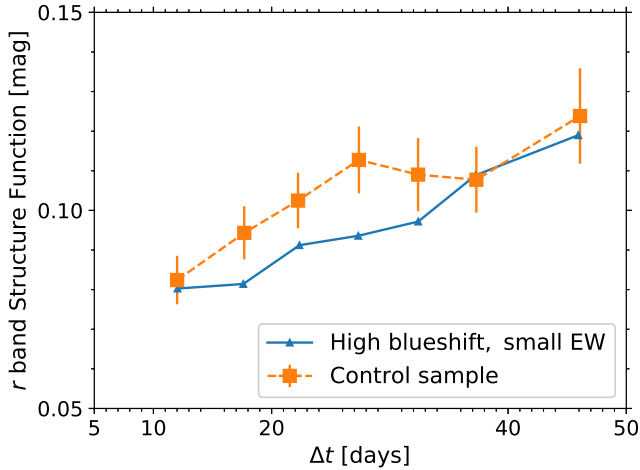}
\caption{The $r$-band variability amplitude as a function of the rest-frame time interval, 
$\Delta t$. Sample A sources (the ``high-blueshift, small-EW'' sources) are less variable, 
indicating higher Eddington ratios. As reported by \cite{Sun2015}, the variability estimation 
on timescales $<10$ days are biased; we therefore only consider quasar variability on longer 
timescales. }
\label{fig:var_cont}
\end{figure}

We then calculated the $r$-band intrinsic variability\footnote{We chose this band because the corresponding 
wavelength is around $\sim 6000\ \rm{\AA}$, which has the smallest spectrophotometric uncertainty \citep{Sun2015}.} 
for sample A and the corresponding control sample. We first calculated the 
synthetic flux in the $r$ band by convolving the $r$-band bandpass with the spectra. As a second step, we 
calculated the structure function \citep[we adopted the IQR estimator; see e.g.,][]{Sun2015} for sample A and 
the control sample. Our results are presented in Figure~\ref{fig:var_cont}. It is 
clear that sources in sample A are intrinsically less variable. Therefore, our results disfavor the orientation 
scenario, but support the idea that high-blueshift sources often be more active (i.e., have higher Eddington 
ratios). \cite{Luo2015} explored the X-ray properties of PHL 1811 analogs and weak-line quasars 
that also show evident \CIV\ blueshifts. They found that these sources tend to suffer from significant (intrinsic) 
X-ray absorption. They proposed that PHL 1811 analogs and weak-line quasars can be well explained if these 
sources have very large Eddington ratios, which appears in line with our Eddington-ratio scenario.

As the accretion rate increases, the temperature of the accretion disk increases, which 
produces more UV photons. Meanwhile, as revealed by recent radiation magnetohydrodynamic simulations 
\citep[e.g.,][]{Jiang2014a}, the energy dissipation efficiency of the X-ray corona decreases with the accretion 
rate. The X-ray corona will be more efficiently cooled due to inverse Compton scattering of these UV photons, 
i.e., the SED becomes softer (i.e., having larger $\alpha_{\mathrm{ox}}$) with the increasing Eddington 
ratio. In addition, the inner accretion disk is puffed up at high Eddington ratios (e.g., 
$\lambda_{\mathrm{Edd}} \gtrsim 0.3$) due to radiation pressure \citep[e.g.,][]{Abramowicz1988, Wang2003, 
Jiang2014b, Sadowski2014}. The puffed-up disk could act as a ``shielding'' gas \citep[e.g.,][]{Leighly2004, Wu2011, 
Luo2015} that blocks both the X-ray coronal emission and the ionizing continuum, i.e., the SED is expected to 
be softer. Quasars with such softer SEDs (i.e., weaker X-ray emission) can launch strong winds from the 
accretion disk \citep[e.g.,][]{Murray1997, Leighly2004, Richards2011, Chajet2013, Chajet2017, Luo2014}. 

The Eddington-ratio scenario has important implications for RM. According to the SED-evolution 
picture, the radius-optical luminosity relation of the low-blueshift sources is invalid for 
the high-blueshift ones; in those cases, the radius and $M_{\rm{BH}}$ will be overestimated. 
Therefore, for the high-blueshift sources, the Eddington ratios we measured (for the methodology, 
see Section~\ref{sect:emlp}) might be lower limits on the true values. A direct test can be applied 
to our scenario by performing RM campaigns \citep[e.g., SDSS-RM;][]{Shen2015} for the extreme 
blueshift sources, and exploring the radius-optical luminosity relation as a function of the \CIV \ 
blueshift \citep[also see][]{Richards2011}.

\subsection{The connection between the \CIV \ blueshift and quasar properties}
\label{sect:dis3}
This Eddington-ratio scenario can also explain additional observational results in this work. 
According to our scenario, when the Eddington ratio increases, the SED 
becomes softer and the covering factor of the shielding gas to the BLR increases \citep{Luo2015}. 
Hence, we expect the BELs to be weaker. \CIV \ and \HeII\ (or other high-ionization lines) 
would be preferentially reduced with respect to \MgII \ (or other low-ionization lines) as the 
ionization energy of the former is larger. Therefore, the Eddington-ratio scenario can plausibly 
explain our results in Figure~\ref{fig:ew_comp}. 

How do we explain the fact that FWHM and $\sigma$ are both anti-correlated with the blueshifts 
(Figures~\ref{fig:fwhm_vof} \&~\ref{fig:sigma_vof})? This relation is apparently inconsistent with 
previous results from \cite{Denney2012}, who found that the rms spectra of \CIV \ (which are 
assumed to represent the variable emissions) are broader than the mean spectra, indicating 
that the non-variable blueshift component should be narrower than the ``canonical'' (or disk) \CIV \ 
profile. As pointed out by \cite{Barth2015}, rms spectra only evaluate the relative variability amplitude 
as a function of the line-of-sight velocity. It is possible to produce very broad rms spectra (broader 
than the single-epoch disk profile) if the high line-of-sight velocity gas responds more efficiently than 
that of the low line-of-sight velocity gas. However, as high velocity components are generally produced 
in the high-ionization region, it is unlikely that this emission is more sensitive to the 
continuum variations \citep{Korista2004}. The Eddington-ratio scenario could provide a plausible 
explanation for why the ratio of the line width (measured as both FWHM and $\sigma$) of \CIV\ to that 
of \MgII\ increases with the \CIV\ blueshift. The radius of the BLR gas should scale as 
$L_{\mathrm{ion}}^{0.5}$, where $L_{\mathrm{ion}}$ is the ionizing continuum luminosity. The ratio of 
the \CIV \ radius to that of \MgII \ decreases with the increasing Eddington ratio. Therefore, 
$\mathrm{FWHM_{CIV}}/\mathrm{FWHM_{MgII}}$ and $\sigma_{\mathrm{CIV}}/\sigma_{\mathrm{MgII}}$ 
are expected to be correlated with the \CIV\ blueshift. The accretion-disk winds may produce 
singly-peaked \citep[which is due to the radiative-transfer effects, e.g., the escape probability is anisotropic; 
see][]{Murray1997} and boxy (i.e., large values of $D$) \CIV\ profiles since they are generated in the inner 
high-speed regions. Therefore, the line profiles of high-blueshift sources are more boxy (i.e., large values 
of $D$) than those of the low-blueshift ones. It is also conceivable that, as the Eddington 
ratio increases, the radiation pressure plays a more important role in accelerating clouds \citep[especially 
low column density ones; see, e.g.,][]{Marziani2010}. Such clouds could produce blueshifted broad \CIV . 
This mechanism could also be (at least partially) responsible for the observed anti-correlation between 
the line-width ratio and the \CIV\ blueshift. 

A remaining question is why quasars with different Eddington ratios/SEDs can have similar \CIV \ EWs. 
Previous works suggest that the ``Baldwin effect'' might be induced by the tight correlation between 
EWs and the Eddington ratios \citep[e.g.,][]{Bachev2004, Baskin2004, Shemmer2015}. However, the \CIV \ 
or \HeII\ EW measures only the ratio of the product of the $E_{\rm{ion}}\gtrsim 50\ \rm{eV}$ extreme 
UV (EUV) emission and the effective covering factor of the BLR clouds to $L1350$. It is possible that 
either $L1350$ or EUV emission cannot effectively track the disk emission \citep[also see][]{Vasudevan2007} 
and/or the effective covering factor varies among quasars. Therefore, the \CIV \ or \HeII\ EW is not an 
accurate indicator of the Eddington ratio or quasar SED.

\subsection{The evolution of the line-shift variability}
How do we understand the evolution of the line-shift variability as a function of the \CIV\ blueshift? Recall that 
the observed line-shift variations are driven by the ``line-of-sight velocity-dependent reverberation'' (see the 
first paragraph of Section~\ref{sect:result_sp}). According to this scenario, the variability amplitude depends 
on the time-lag difference between the blue and red wings and on the variations of the ionizing continuum. 
The time-lag difference is expected if the BLR gas has significant radial motions \citep[see also][]{Barth2015}. 

The line-shift variability of \MgII\ is more extreme than that of \CIV. At a first glance, this result is not expected 
since the ionization energy of \CIV\ is much larger than that of \MgII, and the variability amplitude of quasar 
continuum emission generally increases with energy. We argue that our result could be explained as follows. 
The distance of the location of the \MgII\ gas should be larger than that of \CIV, suggesting that the \MgII\ gas 
is radially more extended. As a result, the \MgII\ time-lag difference between the blue and red wings is larger 
than that of \CIV, which leads to larger line-shift variability of \MgII.

For \CIV, high-blueshift sources tend to have small line-shift variability. This connection might simply reflect the fact 
that, as mentioned in Section~\ref{sect:dis2}, high-blueshift sources are less variable in terms of quasar continua. 
However, the evolution of VAR.($V_{\rm{shift,MgII,se}}$) along the \CIV\ blueshift is not entirely expected. The 
evolution can only be explained if the time-lag difference of the blue and red wings and/or the ratio of the radial motions 
to the virial motions increases with Eddington ratio. Such a correlation may exist because the ratio of the radiative 
force to the gravitational potential of the SMBH increases with the Eddington ratio \citep[e.g.,][]{SSD}. The radiative 
pressure can help generate non-virial motions in the BLR.

\subsection{Is Eddington ratio the sole factor?}
\label{sect:dis4}
\begin{figure}
\epsscale{1.2}
\plotone{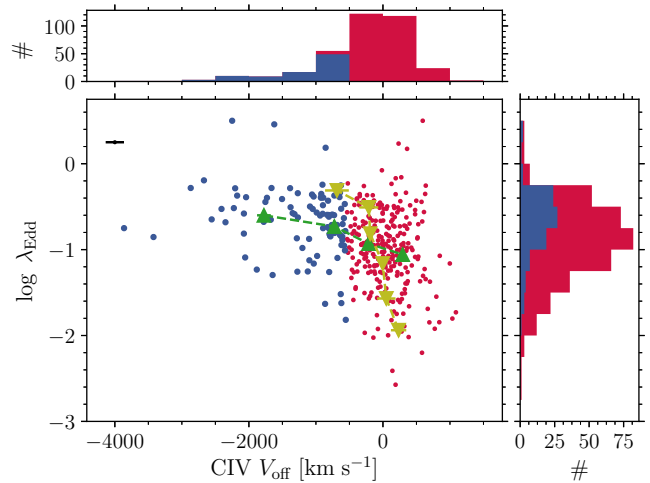}
\caption{Distribution of our sample in the \CIV \ offset velocity-$\lambda_{\rm{Edd}}$ plane. Sources 
with offset velocity $<-550\ \rm{km\ s^{-1}}$ are highlighted by blue colors. The green (yellow) 
triangles represent the mean $\log \lambda_{\rm{Edd}}$ (\CIV \ $V_{\rm{off}}$) in each \CIV \ $V_{\rm{off}}$ 
($\log \lambda_{\rm{Edd}}$) bin. Large blueshift velocities tend to correspond to  large $\lambda_{\rm{Edd}}$. 
However, the reverse is not true. 
}
\label{fig:ledd_vof}
\end{figure}

As we discussed above, our results can be explained if the \CIV\ blueshift is driven by the Eddington ratio 
\citep[which is the driver of the quasar main sequence; see, e.g.,][]{Boroson1992, Sulentic2000b, Shen2014}. 
However, is the Eddington ratio the sole factor to determine the \CIV \ blueshift? Figure~\ref{fig:ledd_vof} 
presents the Eddington ratio as a function of the \CIV \ blueshift. In general, the Eddington ratio and the \CIV\ 
blueshift tend to be anti-correlated. However, some extreme blueshift sources have low Eddington ratios, similar 
to those of no blueshift sources. This 
could simply be caused by the $M_{\rm BH}$ estimation bias discussed above. However, there are 
high Eddington-ratio sources that show almost no blueshift. There are several possibilities for these sources. 
For instance, the high Eddington ratio might only be a necessary but insufficient condition for driving 
accretion-disk winds \citep{Baskin2005, Coatman2016}. Another possibility is that the 
$M_{\mathrm{BH}}$ estimator we adopted in this work (i.e., the \MgII\ single-epoch virial $M_{\mathrm{BH}}$ 
estimator; see Section~\ref{sect:emlp}) has significant intrinsic scatter \citep[$\geq 0.4$ dex; see, 
e.g.,][]{Vestergaard2006} and suffers from considerable Eddington biases \citep{Shen2010}. 
These speculations can be tested by controlling the Eddington ratio and exploring the blueshift as a 
function of quasar luminosity or other properties. Unfortunately, such a study requires unbiased estimation 
of $M_{\rm BH}$ over the entire quasar population, which is presently unavailable. The ongoing RM campaigns, 
such as SDSS-RM \citep{Shen2015} and other multi-object RM campaigns \citep{King2015}, have the potential 
to effectively test our scenario. 

\section{Summary and conclusion}
\label{sect:sum}
We have investigated the \CIV \ blueshift as a function of quasar properties, and constrained the intrinsic 
variability of the \CIV \ blueshift in single-epoch spectra using the $29$ epochs of SDSS-RM spectra. Our 
primary results are as follows. 

\begin{itemize}
\item[1.] We confirmed that the extreme blueshift sources generally have small EWs, while the reverse is 
not true (Figure~\ref{fig:ew_vof}) with our high S/N composite spectra. Other high-ionization emssion lines, 
such as \HeII, also show a blueshift, and the blueshift velocities are correlated with those of \CIV \ 
(Figure~\ref{fig:heii_vof}). Furthermore, the dependence of the \HeII\ blueshift on EWs is similar to that of 
\CIV \ (Figure~\ref{fig:heii_cot}). These results suggest that the blueshift behavior is common for high-ionization 
emission lines (Section~\ref{sect:voffew}). 

\item[2.] Compared with \MgII, \CIV \ is preferentially suppressed for the extreme blueshift 
sources (Figure~\ref{fig:ew_comp}). This result indicates a reduction of the high-energy ionizing continuum 
over the low-energy one (Section~\ref{sect:voffew}). 

\item[3.] $\mathrm{FWHM}_{\mathrm{CIV}}/\mathrm{FWHM}_{\mathrm{MgII}}$ anti-correlates with 
the \CIV \ blueshift (Figure~\ref{fig:fwhm_vof}). Similar relations are also found for 
$\sigma_{\mathrm{CIV}}/\sigma_{\mathrm{MgII}}$, albeit the correlation is not apparent at the extreme blueshift 
(Figure~\ref{fig:sigma_vof}). These relations can be used to make corrections for the \CIV\ $M_{\rm{BH}}$ 
estimators (Section~\ref{sect:width}). 

\item[4.] We also investigated the line-shift variability of \MgII \ (Figure~\ref{fig:mgii_v50_var}) 
and \CIV \ (Figure~\ref{fig:civ_v50_var}). The line-shift variability of \CIV\ and \MgII\ 
are different in terms of variability amplitude and their relation with the \CIV\ blueshift. These differences 
indicate that the structures of \CIV\ and \MgII\ evolve differently as a function of 
the \CIV\ blueshift (Sections~\ref{sect:varmg} \&~\ref{sect:varciv}). We also found that quasar variability can 
slightly enhance the connection between the \CIV \ blueshift and EW (Figure~\ref{fig:var_eff}; Section~\ref{sect:dis1})

\item[5.] We presented the variability of quasar continua as a function of the \CIV \ blueshift. The extreme 
blueshift sources are less variable (Figure~\ref{fig:var_cont}), indicating that 
the high-blueshift sources tend to have high Eddington ratios (Figure~\ref{fig:ledd_v50}, Section~\ref{sect:dis2}).

\item[6.] All these results can be explained if quasar SEDs become softer with increasing Eddington ratios and 
with the presence of X-ray shielding by the inner accretion disk. However, a high Eddington 
ratio might be an insufficient condition for the \CIV\ blueshift (Figure~\ref{fig:ledd_vof}). Future multi-object 
RM experiments can probe our scenario. 
\end{itemize}

\acknowledgments
M.Y.S. thanks K. D. Denney for beneficial discussion. We thank the anonymous referee for his/her helpful comments 
that improved the paper. M.Y.S. and Y.Q.X. acknowledge the support from NSFC-11603022, NSFC-11473026, NSFC-11421303, 
the 973 Program (2015CB857004), the China Postdoctoral Science Foundation (2016M600485), the CAS Frontier Science Key 
Research Program (QYZDJ-SSW-SLH006), and the Fundamental Research Funds for the Central Universities. W.N.B. 
acknowledges support from NSF Grant AST-1516784 and Chandra X-ray Center grant GO6-17083X. 

Funding for SDSS-III has been provided by the Alfred P. Sloan Foundation, the Participating Institutions, the 
National Science Foundation, and the U.S. Department of Energy Office of Science. The SDSS-III web site is 
\url{http://www.sdss3.org/}. 

SDSS-III is managed by the Astrophysical Research Consortium for the Participating Institutions of the SDSS-III 
Collaboration including the University of Arizona, the Brazilian Participation Group, Brookhaven National Laboratory, 
Carnegie Mellon University, University of Florida, the French Participation Group, the German Participation Group, 
Harvard University, the Instituto de Astrofisica de Canarias, the Michigan State/Notre Dame/JINA Participation Group, 
Johns Hopkins University, Lawrence Berkeley National Laboratory, Max Planck Institute for Astrophysics, Max Planck 
Institute for Extraterrestrial Physics, New Mexico State University, New York University, Ohio State University, Pennsylvania 
State University, University of Portsmouth, Princeton University, the Spanish Participation Group, University of Tokyo, 
University of Utah, Vanderbilt University, University of Virginia, University of Washington, and Yale University. 

\software{Astropy \citep{Astropy2013}, Matplotlib \citep{Hunter2007}, Numpy \& Scipy \citep{scipy}}

\end{document}